\documentclass[a4paper,11pt]{article}
\pdfoutput=1 

\usepackage{jcappub} 

\usepackage[T1]{fontenc} 
\usepackage{caption}
\usepackage{subcaption}
\usepackage{mathtools}
\usepackage{xcolor}
\usepackage{journals}
\usepackage[varg]{txfonts}

\renewcommand{\d}{\mathrm{d}}
\newcommand{\D}{\mathrm{d}}
\newcommand{\I}{\mathrm{i}}
\newcommand{\E}{\mathrm{e}}

\title{\boldmath Kinetic field theory: effects of modified gravity theories with screening mechanisms on non-linear cosmic density fluctuations }

\author[a]{A. Oestreicher,}
\author[b]{H. Saxena,}
\author[c]{N. Reinhardt,}
\author[d]{E. Kozlikin,}
\author[e]{J. Dombrowski}
\author[d,1]{and M. Bartelmann\note{Corresponding author.}}

\affiliation[a]{CP3-Origins, University of Southern Denmark, \\ Campusvej 55, DK-5230 Odense M, Denmark}
\affiliation[b]{California Institute of Technology, \\ 1200 East California Boulevard, Pasadena CA 91125, USA}
\affiliation[c]{Institute of Scientific Computing, Heidelberg University, \\ Im Neuenheimer Feld 205, D-69121 Heidelberg, Germany}
\affiliation[d]{Institute for Theoretical Physics, Heidelberg University, \\ Philosophenweg 16, D-69120 Heidelberg, Germany}
\affiliation[e]{Independent Researcher, Stuttgart, Germany}

\emailAdd{alexo@cp3.sdu.dk}
\emailAdd{hsaxena@caltech.edu}
\emailAdd{niklas.reinhardt@iwr.uni-heidelberg.de}
\emailAdd{elena.kozlikin@uni-heidelberg.de}
\emailAdd{Johannes\_Dombrowski@gmx.de}
\emailAdd{bartelmann@uni-heidelberg.de}

\abstract{In a mean-field approximation within Kinetic Field Theory, it is possible to derive an accurate analytic expression for the power spectrum of present-day non-linear cosmic density fluctuations. It depends on the theory of gravity and the cosmological model via the expansion function of the background space-time, the growth factor derived from it, and the gravitational coupling strength, which may deviate from Newton's constant in a manner depending on time and spatial scale. In earlier work \cite{Oestreicher.2023}, we introduced a functional Taylor expansion around general relativity and the cosmological standard model to derive the effects of a wide class of modified-gravity theories on the non-linear power spectrum, assuming that such effects need to be small given the general success of the standard model. Here, we extend this class towards theories with small-scale screening, modeling screening effects by a suitably flexible interpolating function. We compare the Taylor expansion with full mean-field solutions and find good agreement where expected. We find typical relative enhancements of the non-linear power spectrum between a few and a few ten per cent in a broad range of wave numbers between $k\gtrsim0.1-10\,h\,\mathrm{Mpc}^{-1}$, in good qualitative agreement with results of numerical simulations. Taking nDGP gravity as a quantitative example we compare our results to N-body simulations and find percent-level agreement for wavenumbers $k\lesssim 2\,h\,\text{Mpc}^{-1}$, if the scale where screening sets in, $k^*$, is adapted appropriately. This extends the application of our analytic approach to non-linear cosmic structure formation to essentially all classes of modified-gravity theories.}

\begin{document}
\maketitle

\section{Introduction}

The formation and evolution of cosmic structures represent one of the most intriguing puzzles in modern astrophysics and cosmology. Cosmic structures contain valuable information on the expansion history of the background space-time, the nature of dark matter, and the gravitational interaction. As long as the relative density fluctuations of cosmic structures remain small compared to the mean cosmic matter density, their dynamics can be accurately modelled using linear theory. Linear approaches to cosmic structure formation assume that the matter density and the velocity field are governed by the hydrodynamical equations \cite{Bernardeau.2002}. This simplified analytic description is suitable for linear or mildly non-linear, non-relativistic structures with length scales small compared to the Hubble radius.

Wide-area surveys provide insight into cosmic structures across a broad range of scales, characterized by the presence of stars, galaxies, and galaxy clusters. On small scales and at late cosmic times, the density fluctuations become highly non-linear. The trajectories of dark-matter streams can cross at some point in time, leading to a multi-valued velocity field that cannot be captured by hydrodynamics -- the notorious shell-crossing problem. To by-pass this problem, one conventionally relies on $N$-body simulations which follow the trajectories of a large number of tracer particles. However, high-resolution simulations of cosmic structures in the deeply non-linear regime are computationally expensive and do not provide insight into the fundamental processes governing cosmic structure formation.

Kinetic Field Theory (KFT) provides an analytic approach to non-linear cosmic structure formation. It splits the cosmic matter field into classical particles whose trajectories through phase space are governed by the Hamiltonian equations. Thereby avoiding problems due to shell-crossing, since Hamiltonian trajectories cannot cross in phase space. KFT has recently been successfully applied to various aspects of cosmic structure formation (see e.g.\ \cite{Bartelmann.2015, Viermann.2015, Bartelmann.2016, Bartelmann.2016b, Fabis.2018, Bartelmann.2019, Geiss.2021} and \cite{Konrad.2022} for a detailed review). Bartelmann et al.\ \cite{Bartelmann.2021} showed how a mean-field approximation of the KFT interaction operator can be used to calculate a non-linear matter power spectrum that agrees with numerical simulations at the relative level of $\lesssim 5 \,\%$ for wavenumbers $k\lesssim 10\,h\,\text{Mpc}^{-1}$ at redshift $z=0$. This result is the basis for this paper.

Although general relativity (GR) has been tested in numerous ways, a large variety of generalizations and alternatives has been proposed in the last decades to overcome problems such as the necessity of dark matter or dark energy. Changes to the theory of gravity can be included into KFT, and non-linear power spectra can be calculated for them. \cite{Heisenberg.2019} showed generalized Proca theories can be included into the framework of KFT and \cite{Oestreicher.2023} introduced a functional first-order Taylor expansion of the non-linear power spectrum that can be used to study the general response of the power spectrum to a wide class of modified gravity theories. These works however omitted theories which introduce a screening mechanism. Extending the previous work to theories with screening is the purpose of this paper.

The main motivation for screening is the following: Some modified gravity theories aim to avoid the need for dark energy in lieu of a large-distance modification of General Relativity (GR). Such theories require a significant modification of gravity at large scales to explain late-time acceleration. However, they must all tend to GR on small scales to satisfy the tight constraints from observations at Solar-System scales. Screening is then needed to shield gravity at small scales. Modified gravity theories with screening can be effectively described by a density-dependent (i.e.\ environment-dependent) gravitational interaction strength \cite{Hassani.2020}. In this work, we use the effective description of \cite{Hassani.2020} to calculate the non-linear matter density power spectrum of several modified gravity theories with screening effects using KFT. We find generally good agreement with numerical results in the literature on the shape of modifications for the non-linear power spectra. In the exemplary case of nDGP gravity, we compare our results to $N$-body simulations and find percent-level agreement for wavenumbers $k\lesssim 2\,h\,\text{Mpc}^{-1}$.

In summary this paper has two major purposes: Firstly, we wish to extend the analysis of \cite{Oestreicher.2023}, studying the general response of the non-linear power spectrum to changes in the underlying theory of gravity, to theories of gravity involving a screening mechanism. Using an analytic theory enables insight into the origin of changes to structure formation, and provides a framework that can cover a wide range of theories and parameter spaces without having to run computationally expensive simulations. Secondly, extending the formalism to theories involving screening allows us to compare our results to results from numerical simulations in the literature \cite{Cataneo.2019, Hassani.2020}, in order to probe into their agreement with KFT results.

This paper begins in Sect.~\ref{sec:2} with a brief review of the KFT formalism which introduces the main results needed for this work. We introduce the class of modified theories of gravity with screening effects in Sect.~\ref{sec:3} and apply the KFT formalism to them in Sect.~\ref{sec:4}. In Sect.~\ref{sec:5} we adopt the general Taylor approach of \cite{Oestreicher.2023} since it allows to gain insight into the general response of the non-linear power spectrum to a wide class of modified gravity theories.

\section{The KFT mean-field power spectrum}
\label{sec:2}

We study cosmic structures through the density-fluctuation power spectrum, i.e.\ the variance of the Fourier modes of the matter-density fluctuations as a function of wavenumber $k$. KFT is an analytic theory that can be used for this purpose. It describes the statistical properties of many classical particles, subject to Hamiltonian dynamics, both in and out of equilibrium. In doing so, it avoids the shell-crossing problem since Hamiltonian phase-space trajectories cannot cross. The information on the system is encapsulated in a generating functional $Z$, from which statistical properties can be extracted via functional derivatives. Particle interactions are described by an interaction operator $\exp(\I\hat S_\mathrm{I})$ acting on the free generating functional $Z_0$. This operator can either be expanded in a Taylor series, leading to perturbation theory \cite{Bartelmann.2016b, Pixius.2022}, or it can be averaged in a mean-field approach, leading to a mean interaction term $\langle S_\mathrm{I}\rangle$ \cite{Bartelmann.2021}. The power spectrum calculated in the mean-field approximation agrees with numerical results at a level of $\lesssim 5\,\%$ up to wavenumbers $k\lesssim 10\,h\,\text{Mpc}^{-1}$ at redshift zero. This mean-field approach will thus be our starting point. In the following, we briefly review the most important definitions and cite results for the mean-field power spectrum. Full derivations and further discussion can be found in \cite{Bartelmann.2021, Konrad.2022}.

On the expanding spatial background of a Friedmann-Lemaître model universe with scale factor $a$, we introduce comoving coordinates $\Vec{q}$ and use $t = D_+ - 1$ as a new time coordinate, where $D_+$ is the linear growth factor (normalized to unity initially), i.e.\ the solution of the linearized growth equation. Throughout, we assume that the background is unaffected by density fluctuations. The Hamiltonian for the particles is then
\begin{equation}
	H = \frac{\Vec{p}^{\,2}}{2m} + m\varphi\;,
\end{equation}
where the potential $\varphi$ satisfies the Poisson equation
\begin{equation}
	\nabla^2 \varphi = A_{\varphi} \delta\;,\quad
  A_\varphi := \frac{3a}{2m^2}\Omega_\mathrm{m}^\mathrm{(i)}G\;,
	\label{eq:Aphi}
\end{equation}
and $\delta$ is the density contrast relative to the mean background density. $G$ is the gravitational coupling strength. From here on, we set the factor $\Omega_\mathrm{m}^\mathrm{(i)}$ in $A_\varphi$ to unity, since we choose the initial time well within the early matter-dominated era. We also define the effective time-dependent particle mass $m(t) = a^3(t)E(t)\d t/\d a$, where $E(a) := H(a)/H_\mathrm{i}$ is the Hubble function in units of its initial value $H_\mathrm{i}$ at $t_\mathrm{i}$. For convenience, we normalize the scale factor $a$ such that it is unity initially, thereby setting $m(t)$ initially to unity as well. 

The Hamiltonian equations imply the trajectories
\begin{equation}
	q(t) = q^\mathrm{(i)} + g_\mathrm{H}(t,0)p^\mathrm{(i)}-
	\int_0^t\D t'g_\mathrm{H}(t,t')m\nabla\varphi
\end{equation}
through configuration space, where $g_\mathrm{H}(t,t') = \int_{t'}^t\D t''/m(t'')$ is the Hamiltonian Green's function, or propagator. Equivalently, we can study the trajectories 
\begin{equation}
	q(t) = q^\mathrm{(i)} + tp^\mathrm{(i)} -
	\int_0^t\D t'g_\mathrm{H}(t,t')m\nabla\phi\;,
\end{equation}
responding to the effective potential 
\begin{equation}
	\phi = \varphi + A_\varphi D_+ \psi\;,
\end{equation}
with the initial velocity potential $\psi$ defined via $\nabla \psi = p^\mathrm{(i)}$. Since $\psi$ needs to satisfy the continuity equation $\nabla^2 \psi = -\delta^\mathrm{(i)}$, the potential $\phi$ obeys the Poisson equation
\begin{equation}
	\nabla^2\phi = A_\varphi\left(
	\delta - \delta^\mathrm{(lin)}
	\right)\;,
\end{equation}
that is, $\phi$ is sourced by the non-linear density fluctuations only. This demonstrates why introducing new fiducial trajectories and an effective potential acting on them is useful: While the density contrast remains linear, the particles follow the fiducial trajectories $q(t)\approx q^\mathrm{(i)}+t p^\mathrm{(i)}$. Once the linear description of structure growth becomes insufficient at late times and on small-scales, the true particles will increasingly deviate from these trajectories. In Fourier space, the effective potential can be written as
\begin{equation}
	\Tilde{\phi} = -\frac{\tilde A_\varphi}{k^2}\left(
	\Tilde{\delta} - \Tilde{\delta}^\mathrm{(lin)}
	\right)\;, 
\end{equation}
where now $A_\varphi$ also has to be Fourier-transformed because $A_\varphi$ depends on the gravitational coupling strength, which depends on scale for the gravity theories we consider in this work. Using $\Tilde{\phi} = n\Tilde{\delta}\Tilde{v}$, i.e.\ that the potential is a convolution of the fluctuation $\delta n$ of the mean particle number $n$ and the one-particle potential $v$, the one-particle potential can be shown to obey
\begin{equation}
	\Tilde{v} = -\frac{\Tilde A_\varphi}{n(k_0^2 + k^2)}\;, 
	\label{eq:potential}
\end{equation}
which is of Yukawa rather than Newtonian form. The parameter $k_0$ describes the scale where non-linear structure growth sets in. 

The mean-field approximation within KFT allows to express the non-linear, density-fluctuation power spectrum as
\begin{equation}
	P_{\delta}^\mathrm{(nl)}(k,t) \approx
	\E^{\langle S_\mathrm{I} \rangle (k,t)}P_\delta^\mathrm{(lin)}(k,t)\;,
\end{equation}
where the scale- and time-dependent, mean-field averaged interaction term $S_\mathrm{I}$ is
\begin{equation}
	\langle S_\mathrm{I}\rangle (k,t) = -2\I n \vec k\cdot\int_0^t\D t'\, g_\mathrm{H}(t,t')m
	\left(\widetilde{\nabla v}\ast \Bar{P}^\mathrm{(lin)}_\delta\right)(\vec k,t')\;.
	\label{eq:SI0}
\end{equation}
Here $\bar P^\mathrm{(lin)}_{\delta}$ is the damped linear density-fluctuation power spectrum defined as
\begin{equation}
	\bar P^\mathrm{(lin)}_\delta(k,t') = \left(1+Q_\mathrm{D}\right)^{-1}
	D_+^2P^\mathrm{(i)}_\delta(k)\;,
	\label{eq:6}
\end{equation}
with the initial density-fluctuation power spectrum $P^\mathrm{(i)}_\delta$ and the damping term
\begin{equation}
	Q_\mathrm{D} = k^2\lambda^2(t)\;, \quad
	\lambda(t) \approx \frac{t}{1+\sqrt{t/\tau}}\sigma_1\;.
	\label{eq:7}
\end{equation}
Equation \eqref{eq:7} introduces the damping scale $\lambda$, depending on the parameter $\tau$ and the first moment of the density-fluctuation power spectrum $\sigma_1$, with
\begin{equation}
	\sigma_n^2:= \frac{1}{2\pi^2}\int_0^{\infty}\D k\, k^{2n-2}P_{\delta}^\mathrm{(i)}(k,t')\;.
\end{equation}
Returning to Eq.~\eqref{eq:SI0}, inserting Eqs.\ \eqref{eq:potential} and \eqref{eq:Aphi} and evaluating the convolution we find
\begin{equation}
	\langle S_\mathrm{I}\rangle (k,t) =
	3 \int_0^t\D t'\,g_\mathrm{H}(t,t')\frac{a(t')}{m(t')} \vec k \cdot \int_{k'}
	G\left(t',|\vec k -\vec k'|\right)\frac{(\vec k-\vec k')}{k_0^2+(\vec k-\vec k')^2}
	\bar P_{\delta}^{(\mathrm{lin})}(k',t')\;, 
\end{equation}
where $\int_{k} := \int\D^3k/(2\pi)^3$. We have anticipated here that the gravitational coupling $G$ will depend on scale and must therefore be included into the convolution. 

We will later be interested in evaluating the power spectrum for a given redshift $z$ or scale factor $a$. Rewriting the spectrum with $a$ as the independent variable and inserting the definition of the particle mass $m$, we find 
\begin{equation}
	P_{\delta}^\mathrm{(nl)}(k,a) \approx
	\E^{\langle S_\mathrm{I} \rangle (k,a)}P_\delta^\mathrm{(lin)}(k,a)\;,
	\label{eq:P_nl}
\end{equation}
with the interaction term
\begin{equation}
	\langle S_\mathrm{I}\rangle (k,a) =
	3 \int_{a_\mathrm{min}}^a\D a'\,\frac{g_\mathrm{H}(a,a')}{a'^2E(a')} \vec k \cdot \int_{k'}
	G\left(a',|\vec k -\vec k'|\right)\frac{(\vec k-\vec k')}{k_0^2+(\vec k-\vec k')^2}
	\bar P_{\delta}^{(\mathrm{lin})}(k',a')
	\label{eq:SI}
\end{equation}
and the Hamiltonian propagator
\begin{equation}
	g_\mathrm{H}(a,a')= \int_{a'}^a\frac{\D\bar a}{\bar a^3E(\bar a)}\;.
\end{equation}
Evaluating \eqref{eq:P_nl} for the non-linear power spectrum is the essential purpose of this paper.

As \cite{Cataneo.2019, Hassani.2020}, whose results we will later compare to, we choose a flat $\Lambda$CDM cosmology with parameters
$\Omega_{c}h^2= 0.1198,\; \Omega_{b}h^2= 0.02225,\; H_0 = 100h = 68\;\mathrm{km}\;\mathrm{s}^{-1}\mathrm{Mpc}^{-1},\; \sigma_8 = 0.8$ and $n_s= 0.9645$. Our initial power spectrum is a simple primordial spectrum with power law index $n_s$ and the transfer function from \cite{Bardeen.1986} for cold dark matter. We note that \cite{Cataneo.2019} set the initial conditions for their simulations using a initial power spectrum created with \texttt{CAMB} \cite{Lewis.2000}, and \cite{Hassani.2020} set the initial conditions using a power spectrum generated with \texttt{CLASS} \cite{Blas.2011}, which both include Baryonic Acoustic Oscillations (BAOs), which are not included in our spectrum. Since we compare relative differences $P_\mathrm{MG}/P_\mathrm{GR}-1$, this minor difference is not relevant for our paper.

We have introduced two parameters in this section, the non-linear scale $k_0$ and the parameter $\tau$ originating from an effective viscosity $\nu$ that is related to damping. These parameters can either be determined within KFT or matched to give optimal agreement with numerical results. We choose optimized parameters here and keep them constant. They are set by matching our KFT mean-field result to a non-linear spectrum calculated with the halo-model approach by \cite{Mead.2015}. We reiterate that we only perform this matching once for the GR results and then keep these parameters constant throughout the paper. While this is an approximation, adapting them would have a negligible effect compared to the changes caused by modified gravity theories; cf. the discussion in \cite{Oestreicher.2023}. More detail on these parameters can be found in \cite{Bartelmann.2021}.

\section{Parametrization of screening effects in a scale-dependent effective $G$}
\label{sec:3}

Motivated by the variety of screening mechanisms available to scalar-tensor theories, \cite{Lombriser.2016} proposed that a modification of gravity can be parametrized as 
\begin{equation}
	\nabla^2 \psi = \kappa^2 \frac{a^2}{2}\rho_\mathrm{m}
	\frac{G_\mathrm{eff}}{G}
	\delta
\end{equation}
with $\kappa^2 = 8\pi G$. The effective, relative gravitational coupling strength $G_\mathrm{eff}/G$ includes screening and linear suppression mechanisms through the parametrization
\begin{equation}
	\frac{G_\mathrm{eff}}{G} =
	A+\sum_i^{N_0}B_i\prod_j^{N_i}\gamma\left(\frac{r}{r_{0ij}},a_{ij},b_{ij}\right) \;,\quad
	\gamma(x,a,b) = bx^a\left[\left(1+x^{-a}\right)^{1/b}-1\right]\;.
	\label{eq:lomb}
\end{equation}
This defines a class of interpolation schemes between regimes of different radial dependences. In the following we discuss the effects of the different parameters in Eq.\,\eqref{eq:lomb}:
\begin{itemize}
	\item \hypertarget{N0}{$N_0$} and \hypertarget{Ni}{$N_i$} allow to combine different screening mechanisms. For the following discussion, we set $N_0=N_i=1$, i.e.\ we choose one particular screening effect.
	\item \hypertarget{A}{$A$} describes the modification of the gravitational coupling in the fully-screened limit $r\to 0$. We set $A=1$ to recover GR on small scales.
	\item The \hypertarget{B}{$B_i$} quantify by how much gravity is enhanced in the fully unscreened limit ($r\to\infty$).
	\item \hypertarget{r0}{$r_0$} describes the screening scale, i.e.\ the scale below which screening sets in.
	\item \hypertarget{a}{$a$} and \hypertarget{b}{$b$} describe the shape and the amplitude of the interpolation between the screened and unscreened regime.
\end{itemize}
We follow \cite{Hassani.2020} and write the gravitational coupling in Fourier space as
\begin{equation}
	\frac{G_\mathrm{eff}(a,k)}{G} = 1 + B
	\gamma\left(\frac{k^*}{k},a,b\right)\;,
	\label{eq:G_eff}
\end{equation}
where $k^*=r_0^{-1}$ is an effective screening wave number. This leaves four free parameters which describe the modification of gravity:
\begin{itemize}
	\item $b=p_1$, the interpolation-rate parameter;
	\item $B=p_2$, the maximal unscreened modification;
	\item $a= \frac{p_1p_3}{p_1-1}$ with $p_3$, which determines the scaling in the screened limit (i.e. $G_{\textrm{eff}}(a,k)/G\sim k^{-p_3}$ for $k\gg k^*$);
	\item $k^*=p_4$, the screening scale.
\end{itemize}
Later, we need the large-scale limit ($k\to 0$) of Eq.\,\eqref{eq:G_eff}, which reads
\begin{equation}
	\frac{G_\mathrm{eff}^{(\mathrm{lin})}}{G} \equiv \lim_{k\rightarrow 0}\frac{G_\mathrm{eff}}{G}=
	\begin{dcases*} 
		A+B & ($a>0$) \\ 
		A & ($a<0$ and $(a-a/b)<0$) \\
		\infty & ($a<0$ and $(a-a/b)>0$)\;.\\
	\end{dcases*}
	\label{eq:G_lim}
\end{equation}
We shall study three modified gravity theories. We introduce them here and fix their parameter sets $\{p_1,p_2,p_3,p_4\}$.

\subsection{DGP Braneworld Gravity}

Dvali-Gabadadze-Porrati (DGP) gravity \cite{Dvali.2000} describes modified gravity by adding a fifth dimension. Four-dimensional Newtonian gravity emerges on a 3-brane of the 5-dimensional Minkowski space. In DGP gravity, the Vainshtein mechanism \cite{Vainshtein.1972} operates when curvature or local densities become large and hides the extra degree of freedom. It was shown that DGP gravity can explain the late-time cosmic acceleration without the need of exotic matter sources such as dark energy \cite{Deffayet.2002}. On subhorizon scales and in the quasistatic limit, the equations of motion read \cite{Koyama.2007}
\begin{equation}
	\label{eq:DGPEOM}
	\nabla^2\phi + \frac{r_\mathrm{c}^2}{3\beta}\left[
	\left(\nabla^2\phi\right)^2-\left(\nabla_i\nabla^j\phi\right)
	\left(\nabla^i\nabla^j\phi\right)
	\right] = \frac{\kappa^2}{3\beta}\,\rho_\mathrm{m}\delta\;.
\end{equation}
The crossover scale $r_\mathrm{c}$ quantifies the transition from $4D$- to $5D$-gravity, and
\begin{equation}
	\label{eq:ampnDGP}
	\beta = 1+2\sigma H r_\mathrm{c} \left(1+\frac{a}{3}\frac{H'}{H}\right)
\end{equation}
with $\sigma = \pm 1$. Here, the prime denotes the derivative w.r.t.\ the scale factor. Positive $\sigma$ represents the normal branch, which we call nDGP. The negative sign represents the self-accelerating branch. Note that the self-accelerating branch suffers from a ghost instability \cite{Koyama.2007b}, while the normal branch is strongly constrained through observational data \cite{Lombriser.2009,Raccanelli.2012}. We focus on the normal branch because we wish to compare our results to the simulation results of \cite{Hassani.2020}.

Solving Eq.~\eqref{eq:DGPEOM} for a spherical top-hat density gives the effective gravitational coupling
\begin{equation}
	\label{eq:GeffnDGP}
	\frac{G_{\textrm{eff}}}{G}=C_1r^3\biggl[\sqrt{1+C_2r^{-3}}-1\biggr]
\end{equation}
with $C_1=2/(3\beta)$, $C_2=r_\mathrm{v}^3$ and the Vainshtein radius $r_\mathrm{v}$. Comparing with Eq.~\eqref{eq:G_eff}, nDGP gravity can be effectively described using the parameters $p_1=2$, $p_2=1/(3\beta)$ and $p_3=3/2$. Ref. \cite{Schmidt.2009} proposes a time-dependent Vainshtein radius $r_\mathrm{v}$. In contrast, we choose $r_\mathrm{v}=1/k^*$ and evaluate power spectra for different time-independent screening scales $k^*$. Figure \ref{fig:ndgpG} shows $G_{\textrm{eff}}/G$ for two crossover scales $r_\mathrm{c}$ and different screening scales $k^*$. For the crossover scale $r_\mathrm{c}$, we take the values from \cite{Hassani.2020}. For the screening scale $k^*$, we choose reasonable values to show the effect of $k^*$ on $G_{\textrm{eff}}/G$.

Figure \ref{fig:ndgpG} shows the enhancement of gravity on large scales, $k\ll 1$. Since $B=p_2\sim 1/\beta$ and $\beta\sim r_\mathrm{c}$, the modification is larger for smaller crossover scales. At $k\approx k^*$ the screening mechanism sets in, reducing $G_{\textrm{eff}}/G$ until the fully screened limit ($G_{\textrm{eff}}/G=1$) is reached for $k\gg 1$. The specific shape of the modification in the intermediate regime ($k\approx k^*$) is determined through the parameters $p_1$ and $p_3$.

\subsection{k-Mouflage}

Kinetic camouflage or k-mouflage \cite{Babichev.2009} describes a class of scalar-tensor theories in which the scalar field `camouflages' itself via self-interactions in sufficiently strong gravitational fields. Gravity in the Solar System is strong enough for the screening to fulfil the GR constraints on small scales. Solving the Klein-Gordon equation for a static spherically-symmetric matter distribution gives \cite{Lombriser.2016}
\begin{align}
	\label{eq:Kmouflage}
	\frac{G_\mathrm{eff}}{G} =
	\frac{2\beta^2}{\kappa_\chi(r)} = \frac{\beta\kappa}{GM/r^2}\sqrt{-2X}\;,
\end{align}
where $\beta$ is the coupling strength of the model, $\chi = X/\mathcal{M}^4 = -\partial^\mu\phi\partial_\mu\phi/2\mathcal{M}^4$, and $\mathcal{M}$ is a model parameter characterising the suppression scale. We follow \cite{Lombriser.2016} and choose the quadratic model $\kappa(\chi) = -1+\chi+\kappa_0\chi^2$ with $\kappa_0<0$. Solving Eq.~\eqref{eq:Kmouflage} for $\chi(r)$ results in
\begin{equation}
	\label{eq:kmouflagex}
	\frac{G_{\mathrm{eff}}}{G}=C_1\frac{1-\biggl(x^2+1+x\sqrt{x^2+2}\biggr)^{1/3}}{x\,\biggl(x^2+1+x\sqrt{x^2+2}\biggr)^{1/6}}
\end{equation}
with $x=C_2r^{-2}$, $C_1=3\sqrt{2}\beta^2$ and $C_2=3\beta\kappa M/(2\pi\mathcal{M}^2)\sqrt{-3K_0}$. Taking the limits $x\ll 1$ and $x\gg 1$ in Eq.~\eqref{eq:kmouflagex} and comparing with Eq.~\eqref{eq:G_eff} gives $p_2 =\beta^2$ and $p_3 = 4/3$. We further choose $b=3$ and leave the effective screening wave number $k^*$ as a free parameter. The parameter $b$ can be calibrated from $r=r_0$. However, to examine the qualitative behaviour it suffices to fix $b$ at a reasonable value (we take $b=3$) and vary $k^*$.

Figure \ref{fig:kmouflageG} shows $G_{\textrm{eff}}/G$ for two coupling strengths $\beta$ (i.e.\ two values of $p_2$) and a variety of screening scales $k^*$. $G_{\textrm{eff}}/G$ behaves similarly as in nDGP: gravity is enhanced on small scales (determined by $p_2$), less in the intermediate regime (determined by $k^*$) and not in the fully screened limit on large scales. The intermediate regime is larger in k-Mouflage (i.e.\ the decrease is slower there) than in nDGP because we choose the interpolation shape parameter $a=p_1p_3/(p_1-1)$ to be smaller in k-Mouflage than in nDGP. The amplitude $p_2$ for k-Mouflage in Fig.~\ref{fig:kmouflageG} is chosen much smaller compared to nDGP in Fig.~\ref{fig:ndgpG} because the amplitude is time-independent in k-Mouflage, but time-dependent in nDGP (see Eq.~\eqref{eq:ampnDGP}) and smaller in the past. To end up with effects on the non-linear power spectrum (which depends on $G_{\textrm{eff}}/G$ at all times $a>a_{\textrm{min}}$) of the same magnitude, we therefore choose the amplitude of the modification quite small in k-Mouflage.

\subsection{Yukawa Suppression}

We finally consider a scalar-tensor theory in Brans-Dicke representation with a constant parameter $\omega$ and a linear potential $U=-m^2(\omega+3)(2\phi)$. The quasistatic scalar field equation reads
\begin{equation}
	\label{eq:Yukawa}
	\nabla^2 \delta\phi - m^2\delta\phi + \frac{\kappa^2}{3+2\omega}\delta\rho_\mathrm{m}\approx0\;.
\end{equation}
Solving Eq.~\eqref{eq:Yukawa} for a spherical top-hat density and comparing the limits with Eq.~\eqref{eq:G_eff} shows $p_2 = 1/(3+2\omega)$ and $p_3 = -2$ (see \cite[Section 3.4]{Lombriser.2016}). We further follow \cite{Hassani.2020}, who argue that $b=3$ represents the interpolation between linear and non-linear regime best. The screening scale $k^*$ is treated as a free parameter. In the case of Yukawa suppression, $p_3<0$ implies a modification of gravity on small scales and the GR-limit on large scales. This is in contrast to the situations discussed up until now where GR was recovered on small scales in order to satisfy Solar-System constraints.

\begin{figure}[tbp]
	\centering
	\begin{subfigure}{\textwidth}
		\centering
		\includegraphics[width=0.5\linewidth]{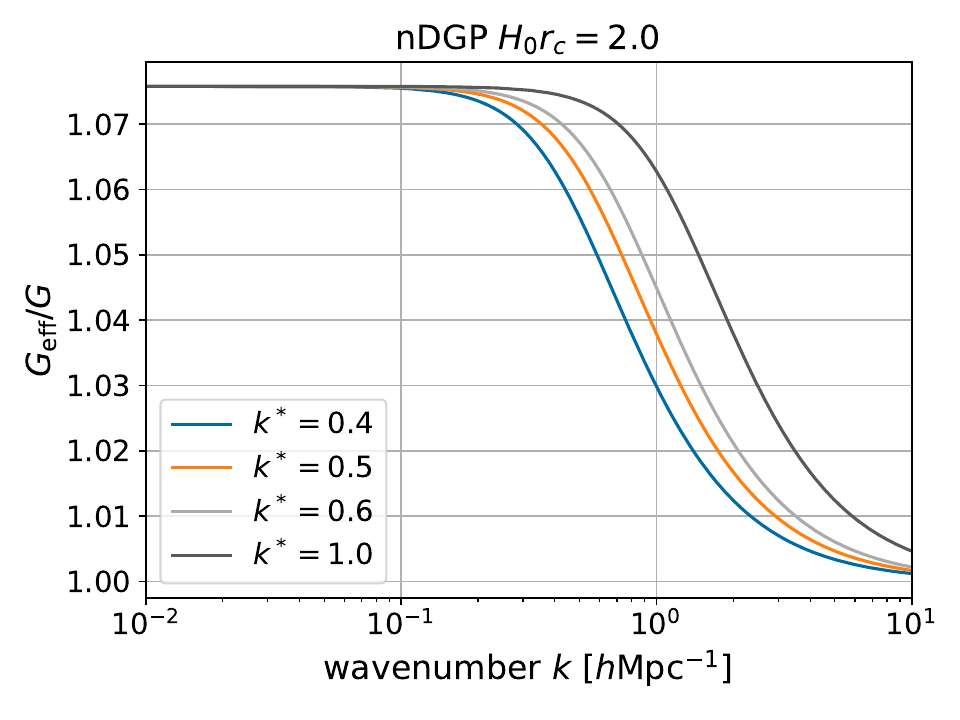}%
		\includegraphics[width=0.5\linewidth]{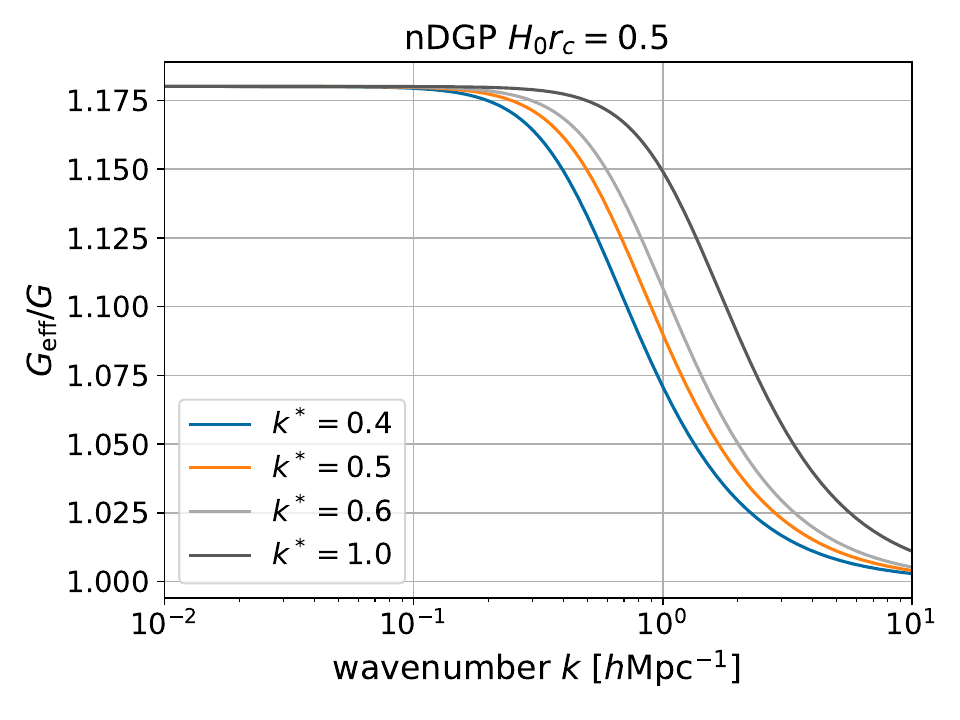}
		\caption{}
		\label{fig:ndgpG}
	\end{subfigure}
	\begin{subfigure}{\textwidth}
		\centering
		\includegraphics[width=0.5\linewidth]{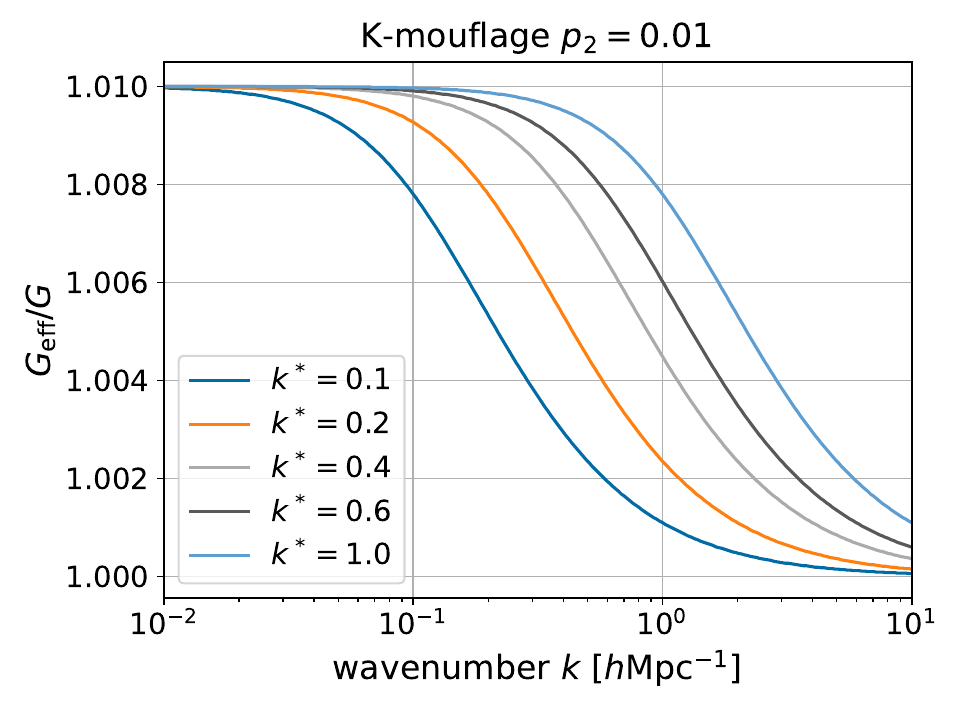}%
		\includegraphics[width=0.5\linewidth]{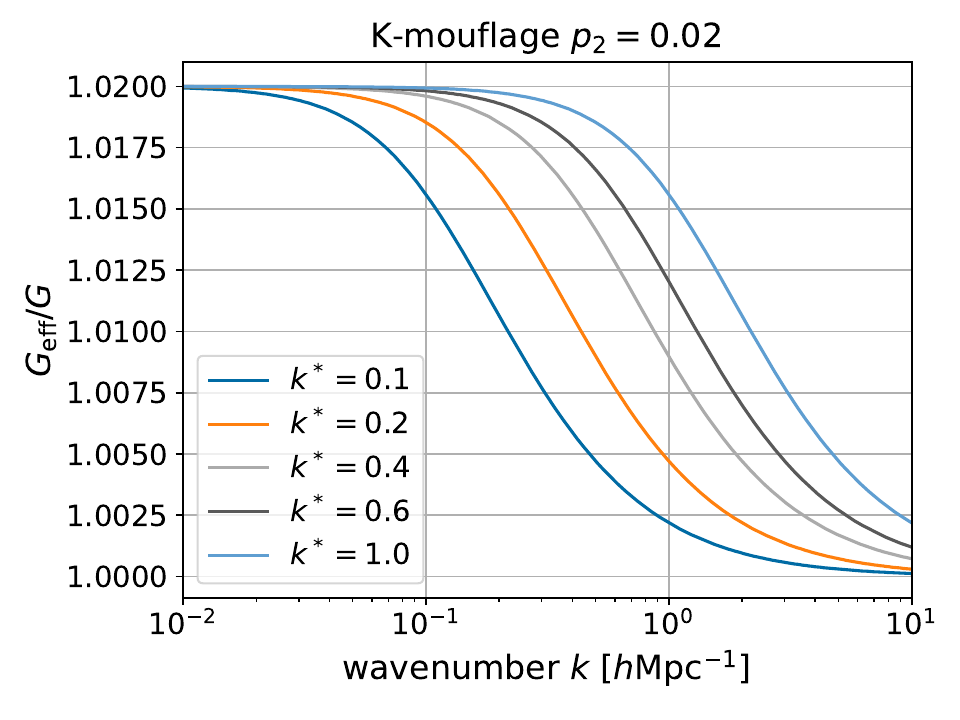}
		\caption{}
		\label{fig:kmouflageG}
	\end{subfigure}
	\begin{subfigure}{\textwidth}
		\centering
		\includegraphics[width=0.5\linewidth]{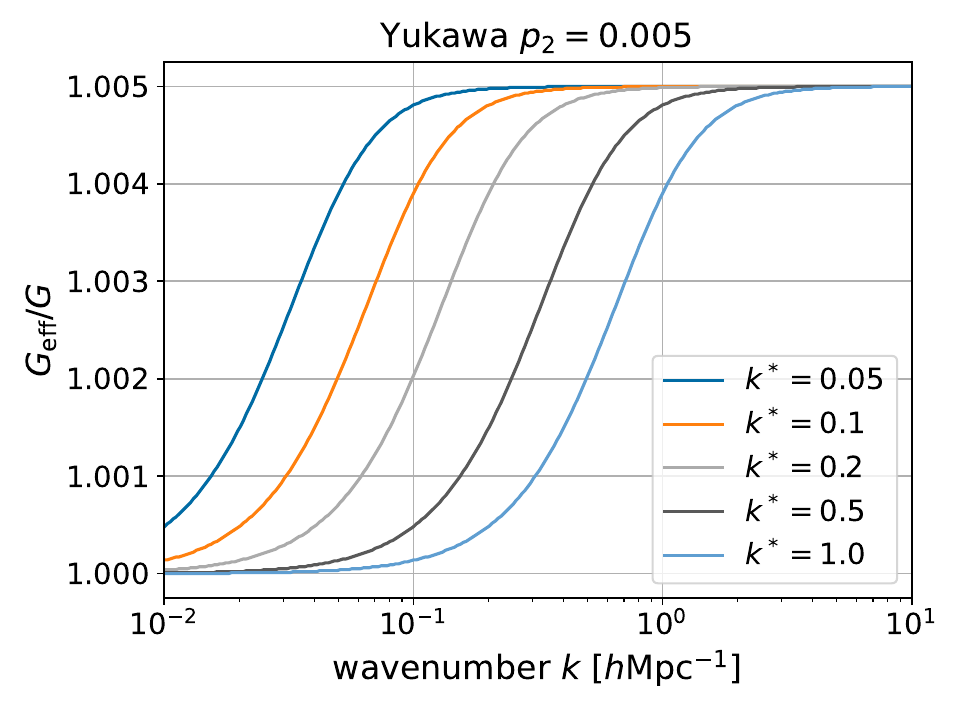}%
		\includegraphics[width=0.5\linewidth]{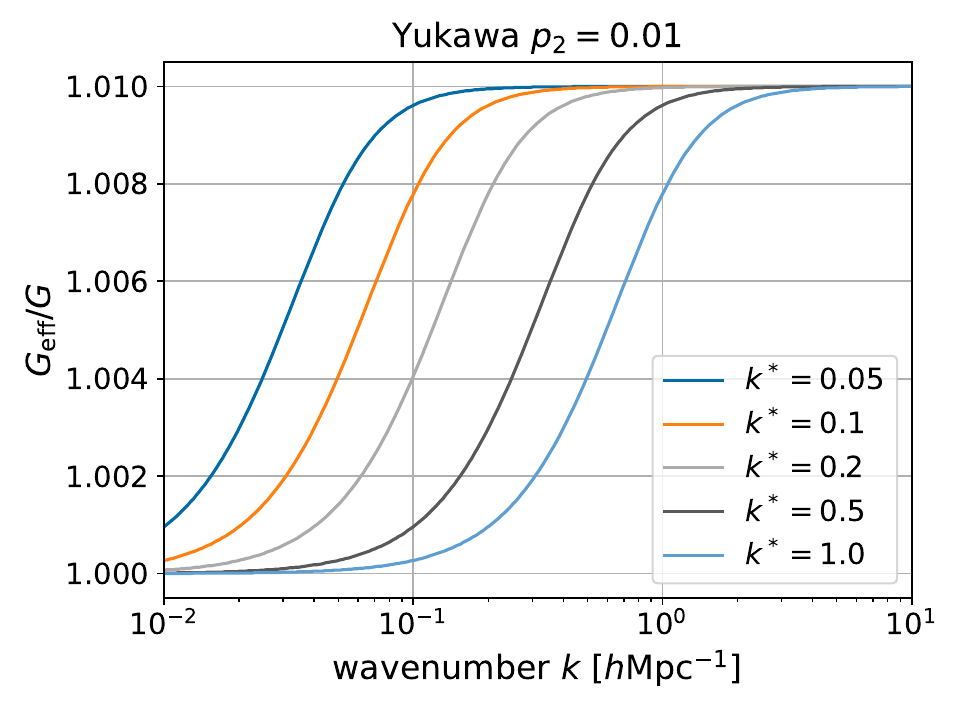}
		\caption{}
		\label{fig:yukawaG}
	\end{subfigure}
	\caption{The relative gravitational coupling strength $G_\mathrm{eff}/G$ evaluated at $z=0$ for the different gravity theories studied in the paper.}
\end{figure}

Figure \ref{fig:yukawaG} shows $G_{\textrm{eff}}/G$ for two amplitudes $p_2$ and different screening scales $k^*$. In contrast to nDGP and k-Mouflage, suppression sets in on large scales, while gravity is enhanced on small scales (depending on $p_2$) with a smooth transition (depending on $k^*$) in between. Since the amplitude $p_2$ is also time-independent in the Yukawa-suppression model, we choose similar values for $p_2$ as for k-Mouflage.

\section{KFT applied to modified gravity with screening mechanisms}
\label{sec:4}

Having adopted a parametrization of the gravitational coupling for modified gravity theories with screening, we now apply KFT to work out the power spectrum of non-linear cosmic density fluctuations in such theories.

\subsection{Formalism}

For this purpose, we need to know three types of input: first, how the gravitational coupling is modified; second, how this influences the linear growth factor $D_+$ and thereby the linear evolution of the power spectrum; and third, how the KFT time coordinate $t$ should be changed. 

We saw earlier in Sect.~\ref{sec:3} how the gravitational coupling needs to be modified in presence of screening. The second input can be taken from the linear growth equation
\begin{align}
	D_+'' + \left(\frac{3}{a} + \frac{E'}{E}
	\right)D_+' - \frac{3\Omega_\mathrm{m} G_\mathrm{eff}(k,a)}{2a^2 G} D_+ = 0\;, 
	\label{eq:lin_growth_eq}
\end{align}
with the matter density parameter defined by
\begin{equation}
	\Omega_\mathrm{m}(a)=\frac{8\pi G}{3H^2(a)}\rho_\mathrm{m}(a)
\end{equation}
as usual. We point out that the growth factor $D_+$ now becomes scale-dependent, hence the linear power spectrum will now evolve as
\begin{equation}
	P_\delta^{(\mathrm{lin})}(k,a)=D_+^2(a,k)P^\mathrm{(i)}_\delta(k)
	\label{eq:Plin_kdep}
\end{equation}
and thus not only change its amplitude but also its shape. While this is not strictly a linear spectrum in the usual sense anymore, since it now contains some of the non-linear effect of the screening mechanism due to the scale dependence of $G$, we keep referring to it as the linear spectrum for the rest of this paper because it results from the linear growth equation. 

The time coordinate $t=D_+-1$ instead of the cosmic time is convenient in KFT. Since this should not depend on scale, we choose the time coordinate as the limit of $D_+-1$ for $k\to0$, where $G_\mathrm{eff}$ and thus also $D_+$ become $k$-independent. This limit has been determined in Eq.~\eqref{eq:G_lim}. We introduce the growth factor $D_+^\mathrm{(lin)}$ calculated with $G_\mathrm{eff}^\mathrm{(lin)}$ as defined there and use the time coordinate $t=D_+^\mathrm{(lin)}-1$.

These three ingredients allow us now to study non-linear density-fluctuation power spectra using the KFT mean-field spectrum introduced in Sect.~\ref{sec:2}.

We begin with Eq.~\eqref{eq:P_nl} for the power spectrum with the interaction term \eqref{eq:SI}. Noting that the second integral in Eq.~\eqref{eq:SI} results from the convolution $\widetilde{\nabla v}\ast \Bar{P}^\mathrm{(lin)}_\delta$ and that the convolution commutes, we rewrite the interaction term as
\begin{equation}
	\langle S_\mathrm{I}\rangle (k,a) = 3 \int_{a_\mathrm{min}}^a\D a'\,\frac{g_\mathrm{H}(a,a')}{a'^2E} \vec k \cdot \int_{k'} G_\mathrm{eff}(a',k')\frac{\vec k'}{k_0^2+k'^2}\bar P_{\delta}^{(\mathrm{lin})}\left(|\vec k - \vec k'|,a'\right)
\end{equation}
which turns out to be numerically advantageous. Introducing spherical polar coordinates in $k$-space and turning $\vec k$ to point along the polar axis, the interaction term becomes
\begin{align}
	\langle S_\mathrm{I}\rangle (k,a) = 3 &\int_{a_\mathrm{min}}^a\D a'\,\frac{g_\mathrm{H}(a,a')}{a'^2E} \int_0^\infty \frac{\D k'\, k'^2}{(2\pi)^2} G_\mathrm{eff}(a',k')\frac{kk'}{k_0^2+k'^2} \nonumber \\
	&\int_{-1}^1\D\mu\, \mu \bar P_{\delta}^{(\mathrm{lin})}\biggl(\sqrt{k^2+k'^2-2kk'\mu},a'\biggr)\;. 
\end{align}
The remaining integral can now be evaluated numerically to yield non-linear power spectra for modified gravity theories with screening. 

One more comment seems to be in order before we turn to studying the specific models introduced in Sect.~\ref{sec:3}. The power-spectrum amplitude is usually set to match observations made today on the largest scales, where the spectrum evolves linearly. As we have seen, the growth factor $D_+$ changes in modified gravity theories. If we fixed the initial power spectrum at some scale factor $a_\mathrm{min}$ to have the same form in standard GR and a modified gravity theory, it would evolve to different amplitudes in the two theories today. Aiming at comparing the present spectrum to observations, we therefore need to set the initial power spectrum for a modified gravity theory with an appropriately changed amplitude in order to end up with a fixed amplitude today. On the largest scales, where the spectrum evolves linearly, the modified gravitational coupling suggested by the parametrisation in Eq.~\eqref{eq:G_eff} becomes $k$-independent. There, the growth factor also becomes $k$-independent, and the amplitude of the initial spectrum needs to be corrected by a factor
\begin{equation}
	\mathcal{A} = \frac{D_{+,\mathrm{GR}}^2(a_\mathrm{norm})}{D_{+,\mathrm{MG}}^{2\mathrm{(lin)}}(a_\mathrm{norm})}\;,
\end{equation}
where $a_\mathrm{norm}$ is the scale factor where the two spectra are set to have the same amplitude. This factor needs to be applied wherever the initial power spectrum appears. This yields
\begin{equation}
	P_{\delta}^\mathrm{(nl)}(k,a) \approx
	\E^{\langle S_\mathrm{I} \rangle (k,a)}\mathcal{A}P_\delta^\mathrm{(lin)}(k,a)
	\label{eq:P_nl_amplitude}
\end{equation}
with the interaction term
\begin{align}
	\langle S_\mathrm{I}\rangle (k,a) = 3 &\int_{a_\mathrm{min}}^a\D a'\,\frac{g_\mathrm{H}(a,a')}{a'^2E} \int_0^\infty \frac{\D k'\, k'^2}{(2\pi)^2} G_\mathrm{eff}(a',k')\frac{kk'}{k_0^2+k'^2} \nonumber \\
	&\int_{-1}^1\D\mu\, \mu \mathcal{A}\bar P_{\delta}^{(\mathrm{lin})}\biggl(\sqrt{k^2+k'^2-2kk'\mu},a'\biggr)
\end{align}
and the damped linear power spectrum is
\begin{equation}
	\bar P^\mathrm{(lin)}_\delta(k,t') = \left(1+\mathcal{A}Q_\mathrm{D}\right)^{-1}
	D_+^2P^\mathrm{(i)}_\delta(k)\;,
\end{equation}
since $Q_\mathrm{D}$ depends on a $k$-space moment of the initial power spectrum. 

\subsection{Results}

\subsubsection{DGP Braneworld Gravity}

\begin{figure}[tbp]
	\begin{subfigure}[]{\textwidth}
		\includegraphics[width=0.5\linewidth]{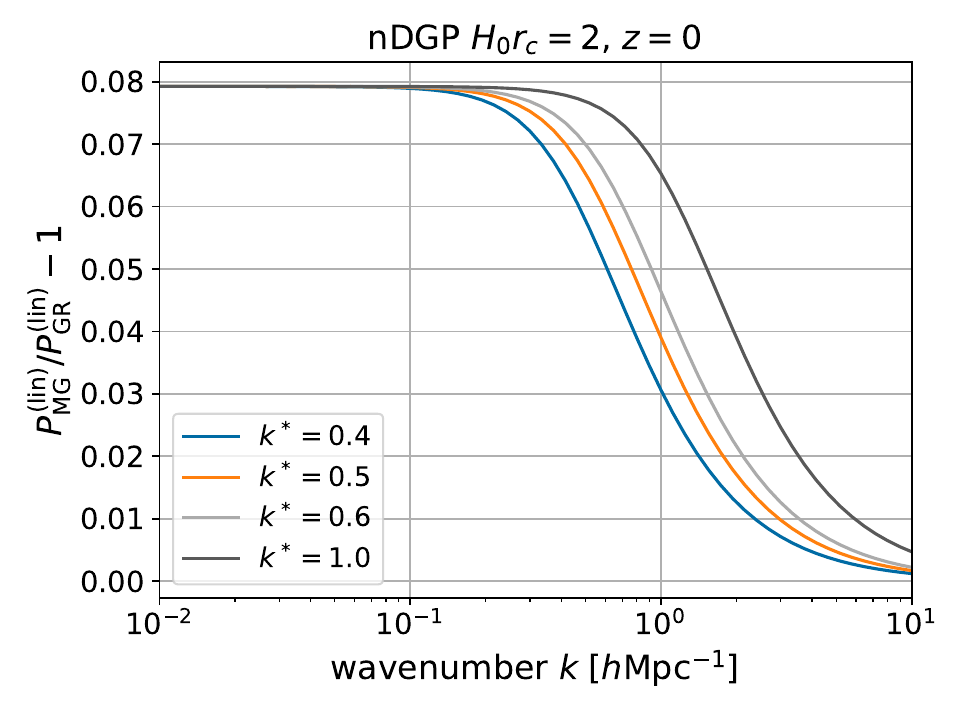}%
		\includegraphics[width=0.5\linewidth]{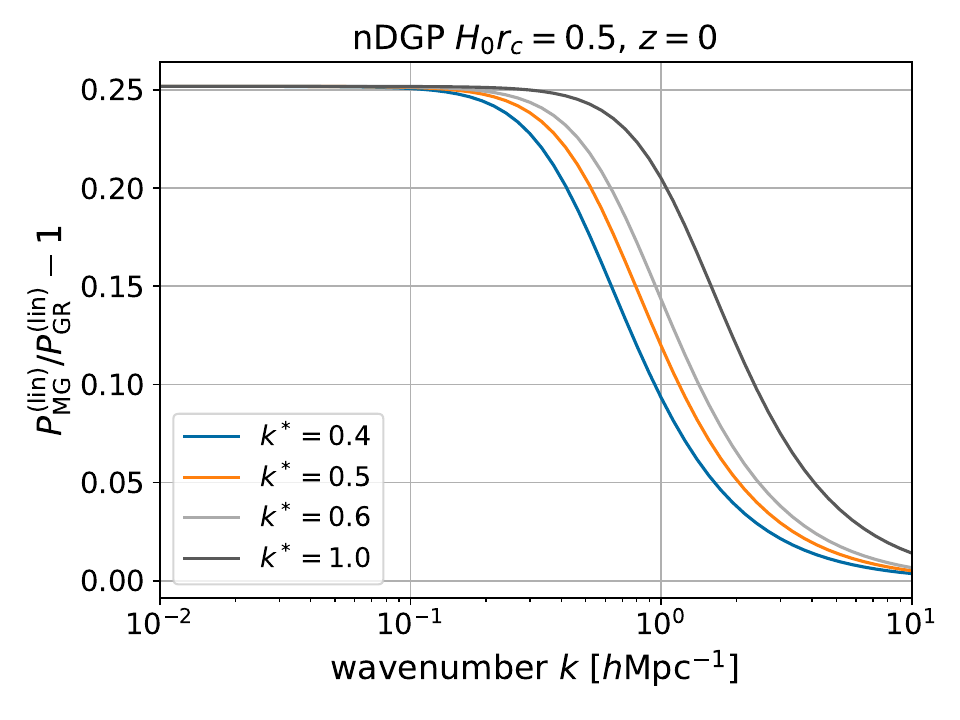}
		\caption{Relative difference between the linear power spectrum (as def. by \eqref{eq:Plin_kdep}) for GR and the nDGP model for two different crossover scales $r_\mathrm{c}$ and a range of different screening wave numbers, evaluated at $z=0$ and normalized at $z\simeq 1000$.}
		\label{fig:ndgp_linear}
	\end{subfigure}
	\begin{subfigure}[]{\textwidth}
		\centering
		\includegraphics[width=0.5\linewidth]{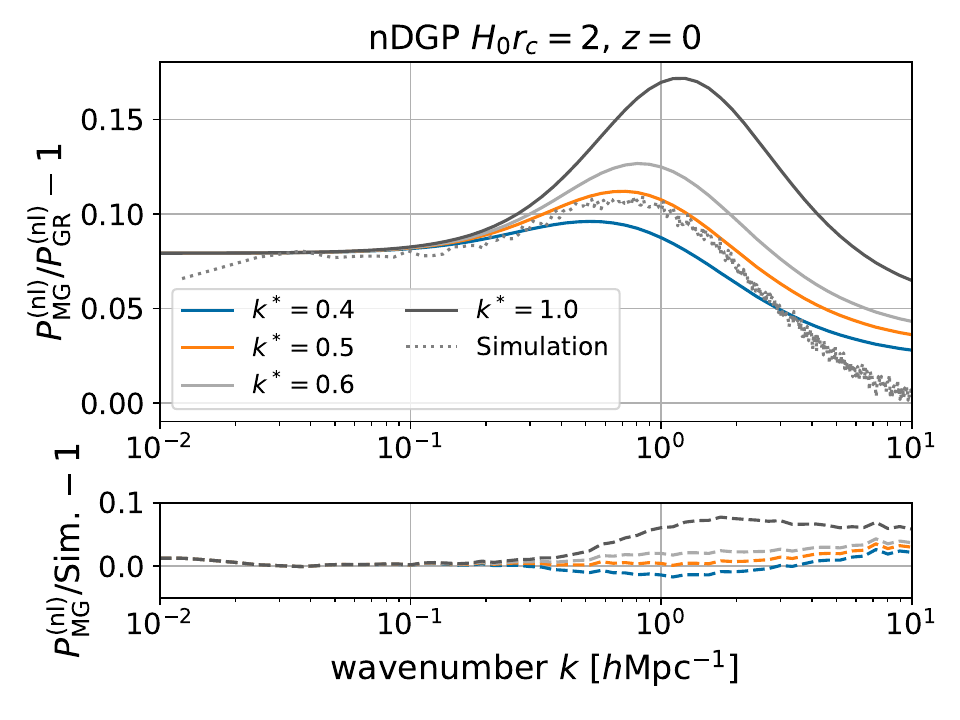}%
		\includegraphics[width=0.5\linewidth]{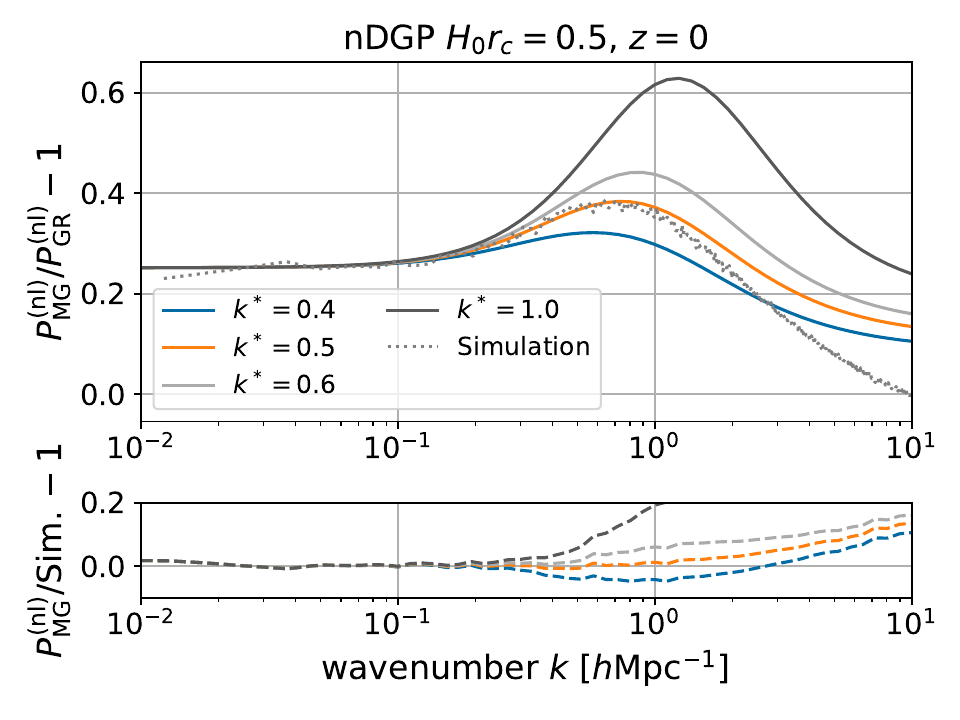}
		\caption{Relative difference between the non-linear power spectrum for GR and the nDGP model for two different crossover scales $r_\mathrm{c}$ and a range of different screening wave numbers, evaluated at $z=0$ and normalized at $z\simeq 1000$. Together with the numerical results from \cite{Cataneo.2019} and the relative difference between our nDGP power spectra and the simulation result.}
		\label{fig:ndgp_nonlinear}
	\end{subfigure}
	\begin{subfigure}[]{\textwidth}
		\centering
		\includegraphics[width=0.5\linewidth]{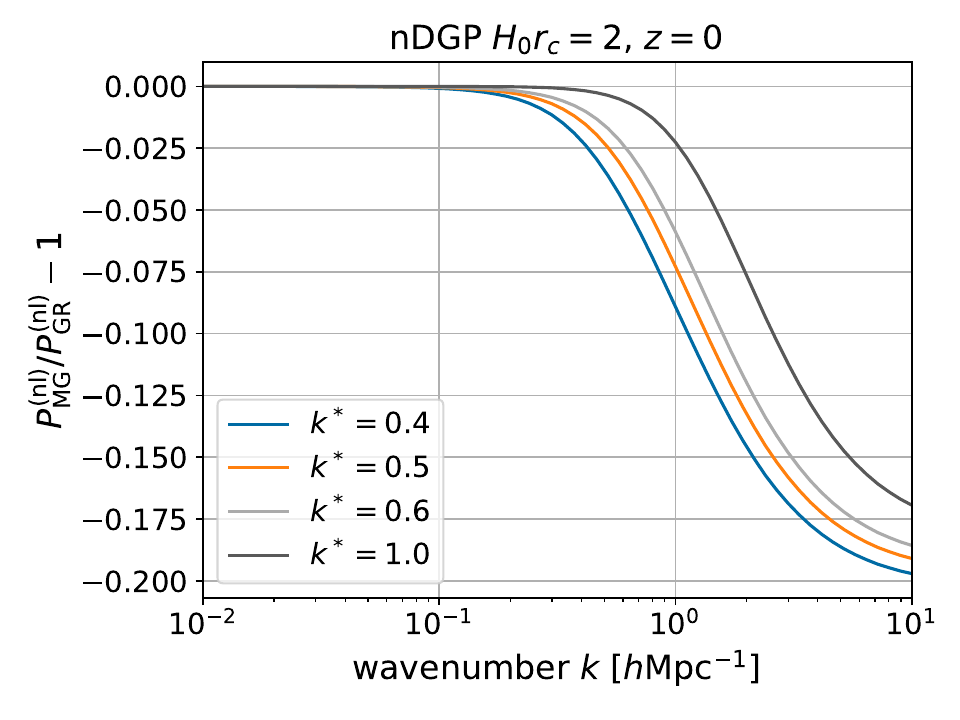}%
		\includegraphics[width=0.5\linewidth]{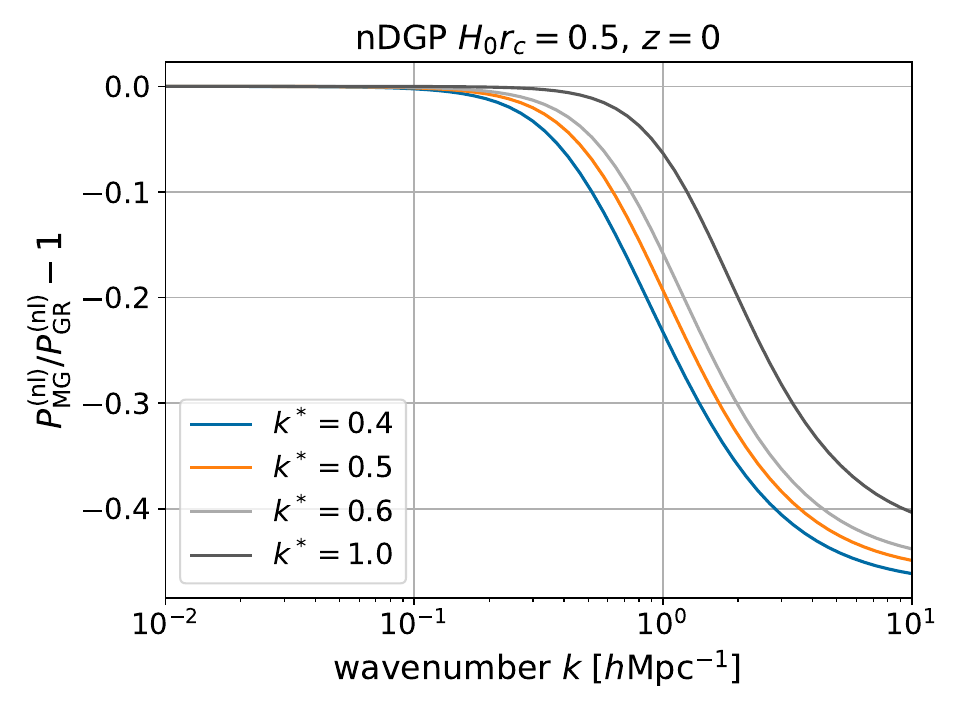}
		\caption{Relative difference between the non-linear power spectrum for GR and the nDGP model for two different crossover scales $r_\mathrm{c}$ and a range of different screening wave numbers, evaluated and normalized at $z=0$.}
		\label{fig:ndgp_nonlinear_an1}
	\end{subfigure}
	\caption{}
\end{figure}

Our first example is the nDGP model introduced in Sect.~\ref{sec:3}. We set the amplitude of the intial spectra for GR and the nDGP model to the same value because we want to compare our results to \cite{Hassani.2020}. While \cite{Hassani.2020} start their evolution at redshift $z=100$, we choose to keep our minimum scale factor at $a_\mathrm{min}=0.001$ corresponding to $z\simeq 1000$. Since the spectra for the nDGP model evolve like GR spectra at such early times, this choice does not affect the results. We first have a look at the linear power spectrum. Figure~\ref{fig:ndgp_linear} shows its relative difference to the GR reference for two choices of $r_\mathrm{c}$ and a range of $k^*$ values evaluated at $z=0$. On the largest scales, $k\to0$, the amplitude of the spectrum changes by a constant factor. For larger $k$, the relative difference decreases, and the spectrum returns to the GR case. The scale where this happens depends on $k^*$, while $r_\mathrm{c}$ determines the magnitude of the change. Lower $r_\mathrm{c}$ implies larger changes to the spectrum. The linear power spectra in Fig.~\ref{fig:ndgp_linear} reflect the shape of the effective gravitational coupling in Fig.~\ref{fig:ndgpG}.

Interactions, as included in Fig.~\ref{fig:ndgp_nonlinear}, increase the relative difference on intermediate scales, creating a peak. On the largest scales, the relative change decreases again and should eventually vanish. The peak location changes with $k^*$, and the transition scale $r_\mathrm{c}$ determines its amplitude.

The authors of \cite{Hassani.2020} calculate non-linear power spectra using $N$-body simulations together with the effective parametrization \eqref{eq:G_eff} of the gravitational interaction. We reproduce the shape of their nDGP results (Figure 3 in \cite{Hassani.2020}) very well. However, our results depend much more strongly on the screening wave number $k^*$. This dependence intuitively seems to make sense since $k^*$ separates screened from unscreened scales. In Fig.~\ref{fig:ndgp_nonlinear}, we include the exact model simulations of \cite{Cataneo.2019} as the dotted lines. For both crossover scales, we achieve the best agreement for $k^*=0.5\,h\,\text{Mpc}^{-1}$ and find percent-level accuracy for $k\lesssim 2.0\,h\,\text{Mpc}^{-1}$. For larger $k$, KFT tends to overestimate the power on non-linear scales w.r.t.\ the $N$-body simulations. 

\begin{figure}[tbp]
	\centering
	\includegraphics[width=0.5\linewidth]{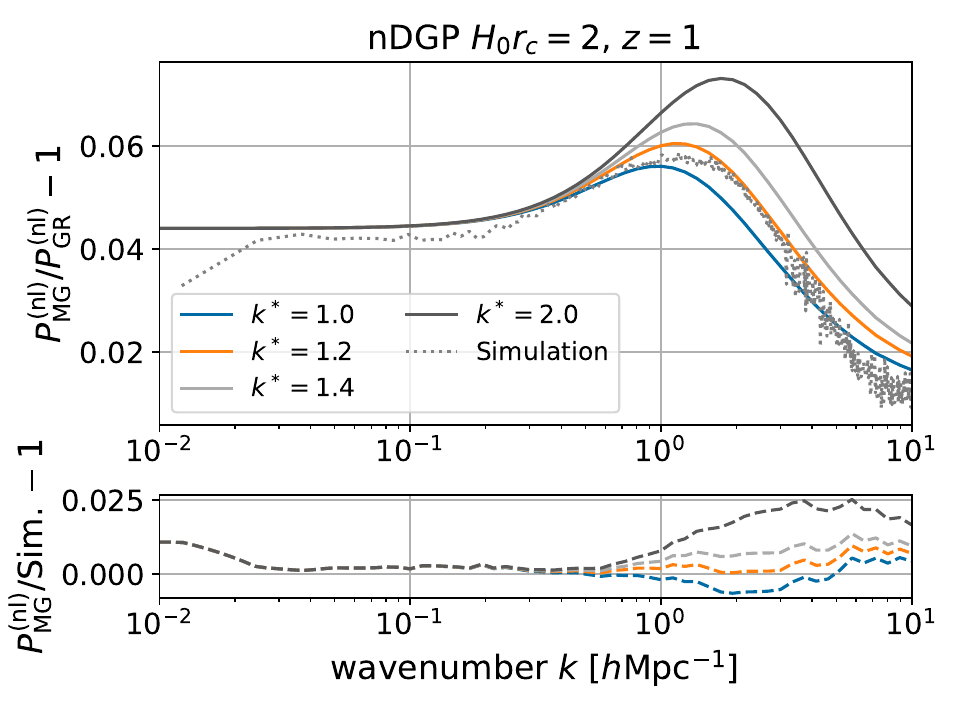}%
	\includegraphics[width=0.5\linewidth]{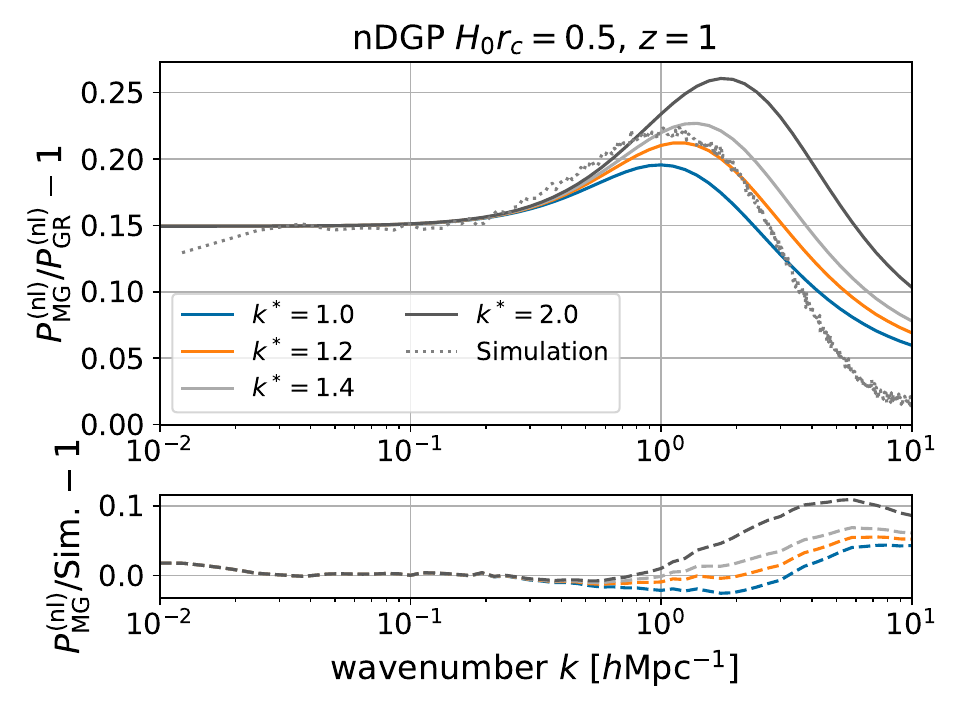}
	\caption{Relative difference between the non-linear power spectrum for GR and the nDGP model for two different crossover scales $r_\mathrm{c}$ and a range of different screening wave numbers, evaluated at $z=1$ and normalized at $z\simeq 1000$. Together with the numerical results from \cite{Cataneo.2019} and the relative difference between our nDGP power spectra and the simulation result.}
	\label{fig:ndgp_nonlinear_a0.5}
\end{figure}

We show the nDGP power spectra evaluated $z = 1$ in Fig.~\ref{fig:ndgp_nonlinear_a0.5}. Once more, KFT with the mean-field approximation reproduces the numerical results of \cite{Cataneo.2019} quite well. Compared to the spectra at $z = 0$, we now find the best agreement for $k^* = 1.2\,h\,\text{Mpc}^{-1}$. This is expected because the numerical results were produced with a time dependent $k^*$.

In both the KFT approach and the approach in \cite{Hassani.2020} a simplified parameterized model was used for modified gravity, where $k^*$ is constant in time for simplicity and is obtained by fit with rigorous simulations. Both approaches can claim that they can reproduce the shape of the power spectrum correctly, but they disagree on the parameter $k^*$. While we find the best agreement for $k^*=0.5\,h\,\text{Mpc}^{-1}$ at $z=0$ and $k^* = 1.2\,h\,\text{Mpc}^{-1}$ at $z=1$ for both values of $r_c$, \cite{Hassani.2020} find the best agreement for $k^*=1.9\,h\,\text{Mpc}^{-1}$ at $z=0$ and $k^* = 2.7\,h\,\text{Mpc}^{-1}$ at $z=1$ for $H_0 r_c=0.5$ and $k^*=0.9\,h\,\text{Mpc}^{-1}$ at $z=0$ and $k^* = 1.3\,h\,\text{Mpc}^{-1}$ at $z=1$ for $H_0 r_c=2$ (see Fig.~3 in \cite{Hassani.2020}). For our results the location of the peak and amplitude is also clearly correlated with $k^*$, both strictly increasing with $k^*$. For \cite{Hassani.2020} there seems to be no such trend, with the peak remaining at roughly the same $k$ independent of $k^*$ and the amplitude not strictly increasing with $k^*$. 

This difference may not be surprising as the oversimplification of assuming a parametrization with constant $k^*$ could in both models lead to different, model-dependent errors. Also, both approaches differ in details of the underlying assumptions that may influence the results. We have assumed the two parameters $k_0, \tau$ to remain unchanged with respect to GR and used a simpler initial spectrum without BAOs. \cite{Hassani.2020} use relatively small simulations (side length $200\;\mathrm{Mpc}/h$) with few particles ($256^3$) and a Gaussian filter to smooth the noisy simulated spectra. Despite these differences, both approaches agree remarkably well with the more accurate results from \cite{Cataneo.2019}, needing only to adapt the fitting parameter $k^*$. 

For comparing our results to observations today, the amplitude of the two spectra should agree on the largest scales, as discussed before. Choosing the amplitude $\mathcal{A}$ in Eq.~\eqref{eq:P_nl_amplitude} accordingly, the relative difference turns negative on the smallest scales, since the nDGP spectrum has to begin at a lower amplitude on all scales and evolves as in GR on the smallest scales; see Fig.~\ref{fig:ndgp_nonlinear_an1}.

\subsubsection{k-Mouflage}

\begin{figure}[tbp]
	\begin{subfigure}[]{\textwidth}
		\includegraphics[width=0.5\linewidth]{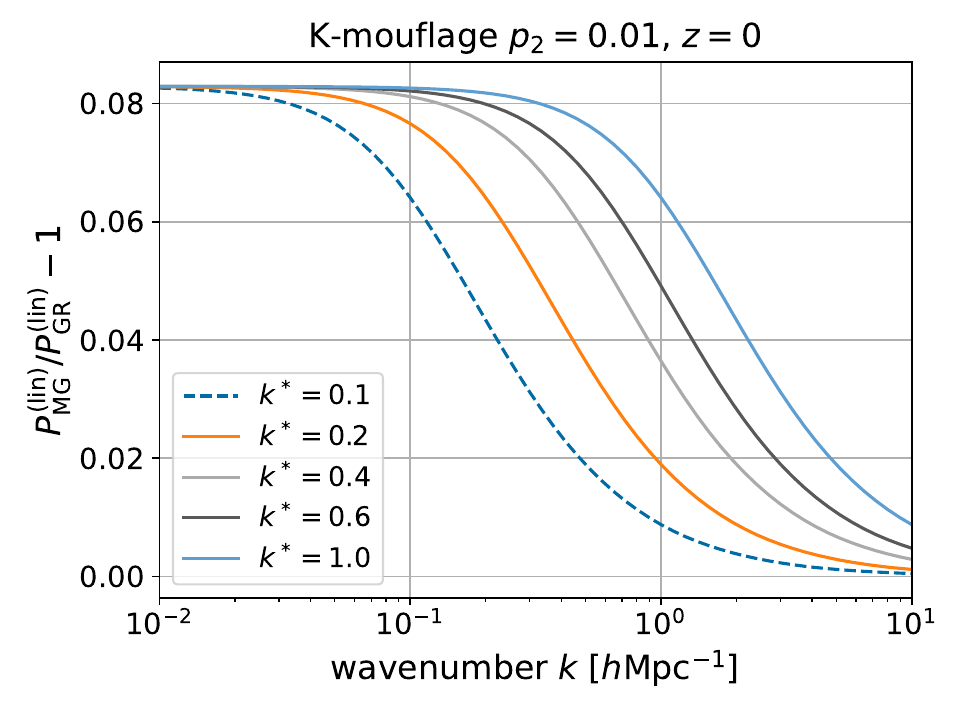}%
		\includegraphics[width=0.5\linewidth]{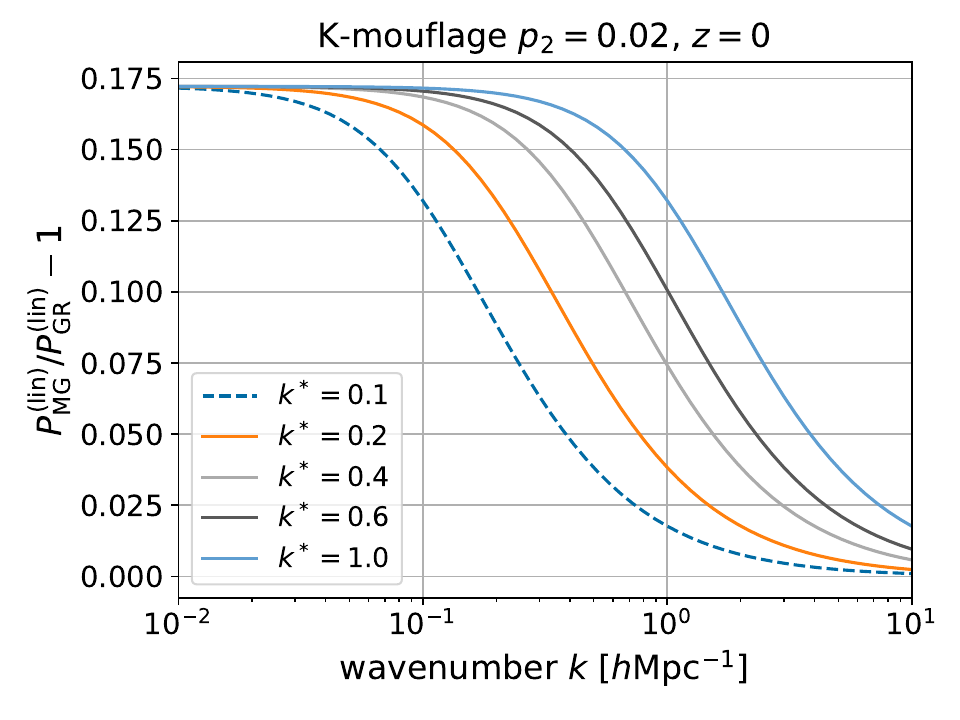}
		\caption{Relative difference between the linear power spectrum (as def. by \eqref{eq:Plin_kdep}) for GR and the K-mouflage model for two different amplitudes $p_2$ and a range of different screening wave numbers, evaluated at $z=0$ and normalized at $z\simeq 1000$.}
		\label{fig:kmouflage_linear}
	\end{subfigure}
	\begin{subfigure}[]{\textwidth}
		\centering
		\includegraphics[width=0.5\linewidth]{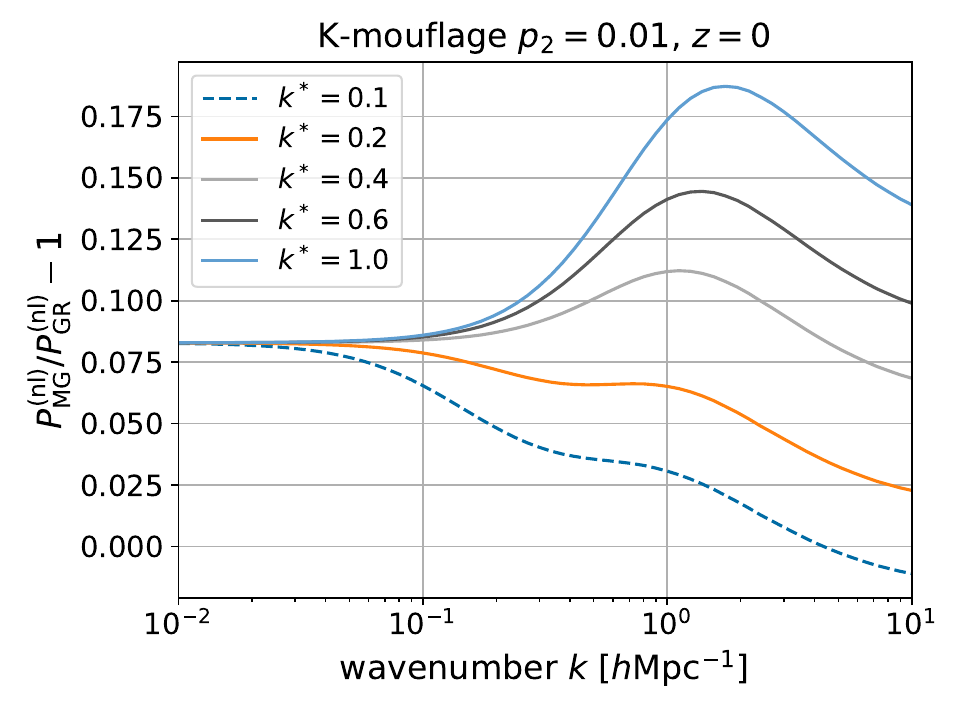}%
		\includegraphics[width=0.5\linewidth]{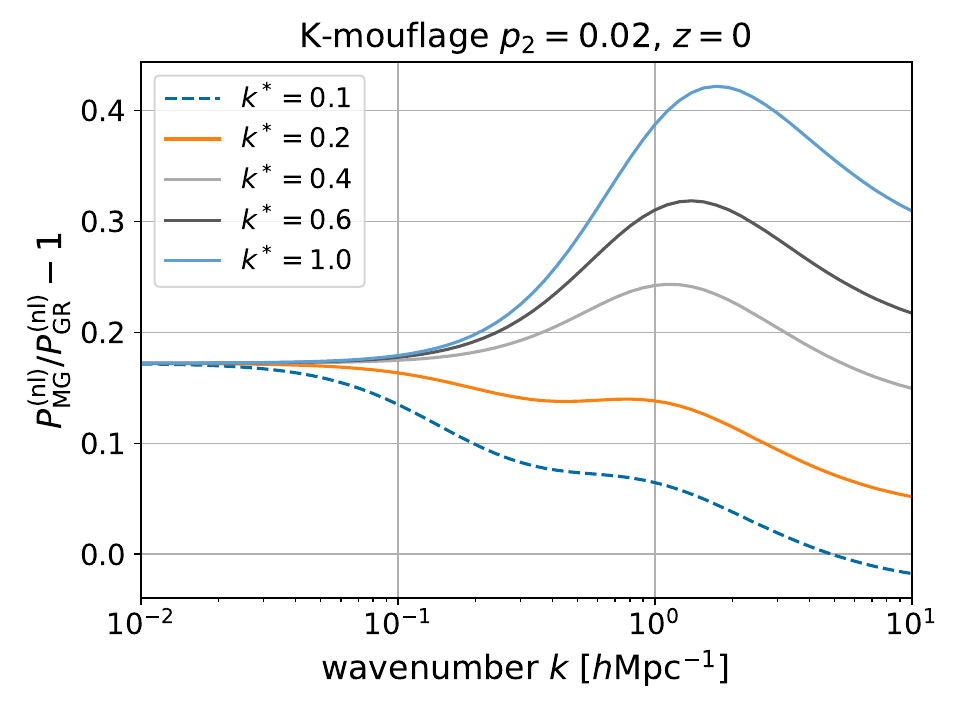}
		\caption{Relative difference between the non-linear power spectrum for GR and the K-mouflage model for two different amplitudes $p_2$ and a range of different screening wave numbers, evaluated at $z=0$ and normalized at $z\simeq 1000$.}
		\label{fig:kmouflage_nonlinear}
	\end{subfigure}
	\begin{subfigure}[]{\textwidth}
		\centering
		\includegraphics[width=0.5\linewidth]{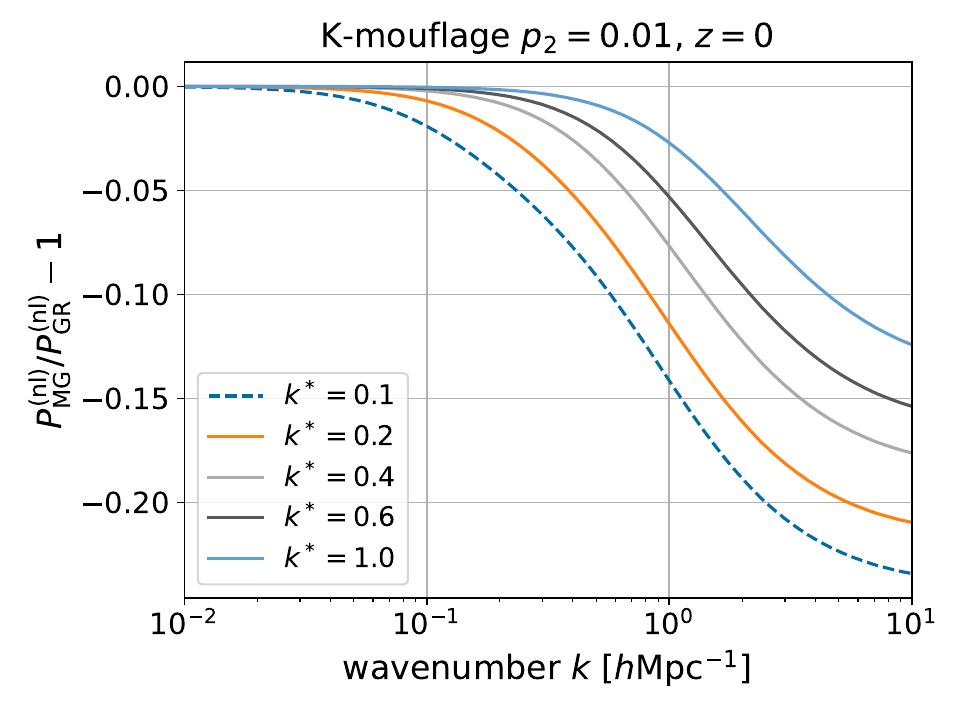}%
		\includegraphics[width=0.5\linewidth]{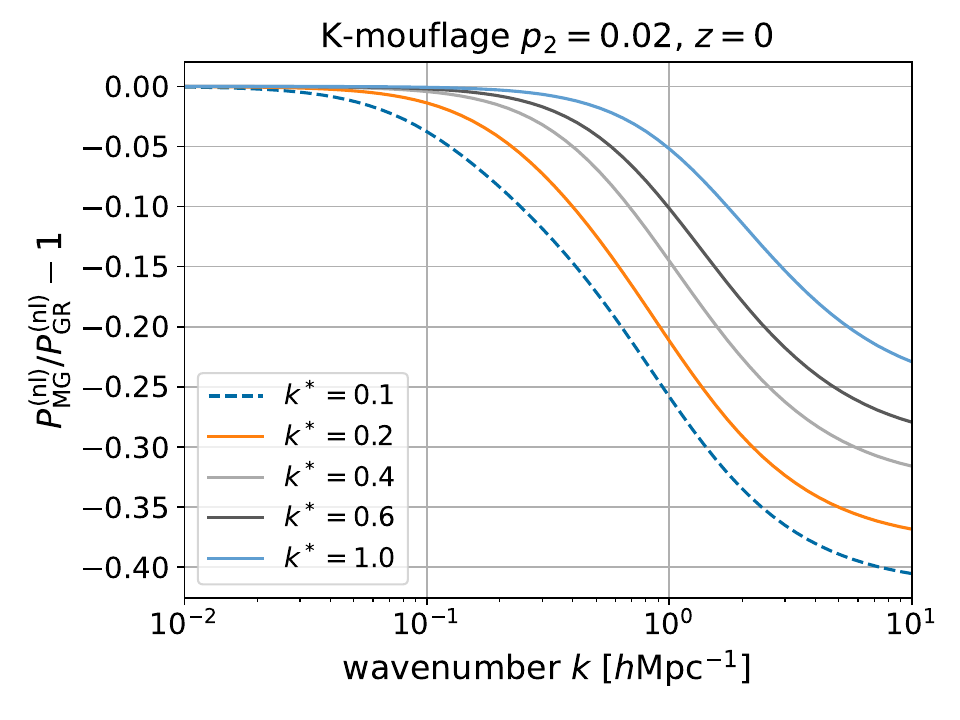}
		\caption{Relative difference between the non-linear power spectrum for GR and the K-mouflage model for two different amplitudes $p_2$ and a range of different screening wave numbers, evaluated and normalized at $z=0$.}
		\label{fig:kmouflage_nonlinear_an1}
	\end{subfigure}
	\caption{}
	\label{fig:kmouflage}
\end{figure}

We now show similar results for the k-Mouflage model introduced in Sect.~\ref{sec:3}, for two choices of $p_2$ and a range of $k^*$ values. The power spectrum changes in a way very similar to nDGP, which was to be expected due to the similar modification of the effective gravitational coupling. The parameter $p_2$ changes the overall amplitude, and the screening wave number $k^*$ sets the scale where screening sets in. We include some curves with very small $k^*$ here. In Fig.~\ref{fig:kmouflage_nonlinear}, which shows the relative difference between non-linear power spectra normalized at $z\simeq 1000$, we see that the relative difference almost returns to zero for $k^*=0.2\,h\,\text{Mpc}^{-1}$ at large $k$, as expected. For $k^*=0.1\,h\,\text{Mpc}^{-1}$, the relative difference turns negative. We suppose that this is an artifact of limitations of the mean-field approximation in the deeply non-linear regime.

\subsubsection{Yukawa Supression}

\begin{figure}[tbp]
	\begin{subfigure}[]{\textwidth}
		\includegraphics[width=0.5\linewidth]{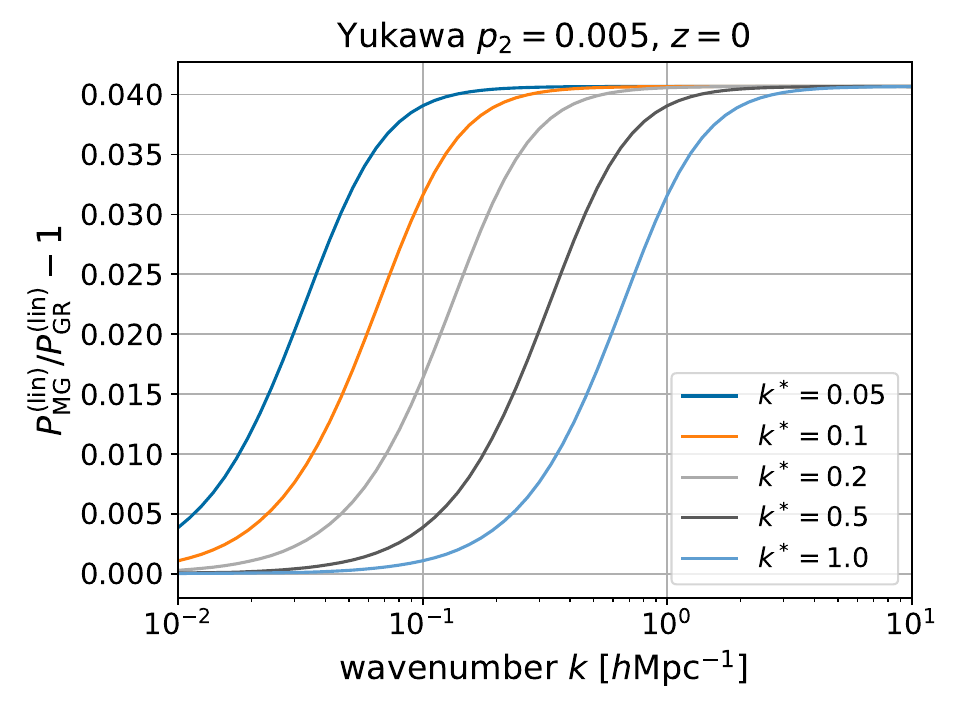}%
		\includegraphics[width=0.5\linewidth]{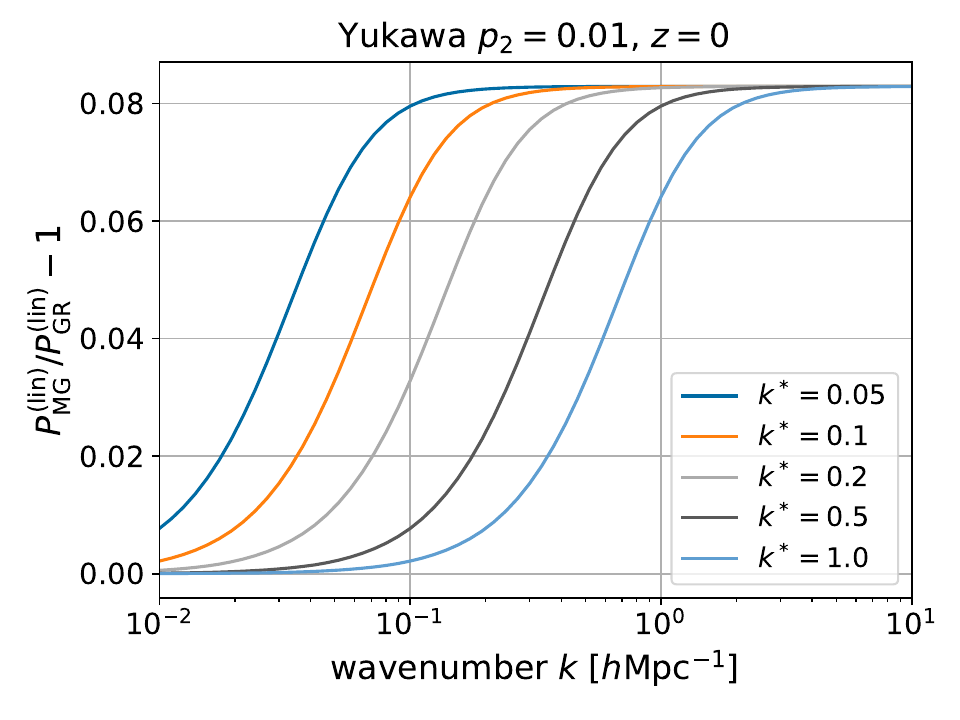}
		\caption{Relative difference between the linear power spectrum (as def. by \eqref{eq:Plin_kdep}) for GR and the Yukawa model for two different amplitudes $p_2$ and a range of different screening wave numbers, evaluated at $z=0$.}
		\label{fig:yukawa_linear}
	\end{subfigure}
	\begin{subfigure}[]{\textwidth}
		\centering
		\includegraphics[width=0.5\linewidth]{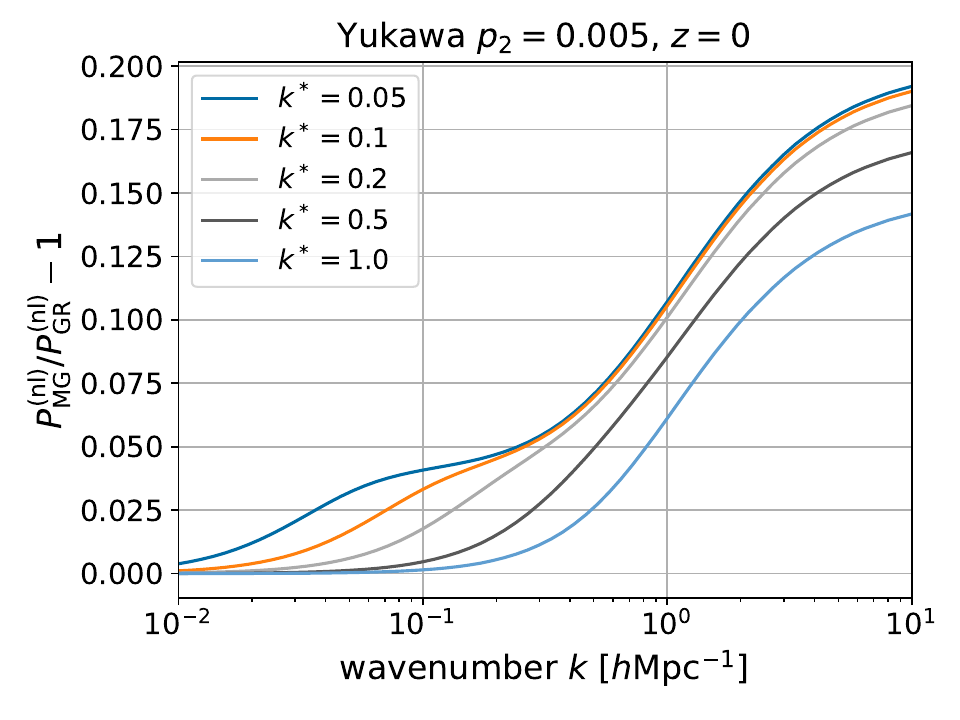}%
		\includegraphics[width=0.5\linewidth]{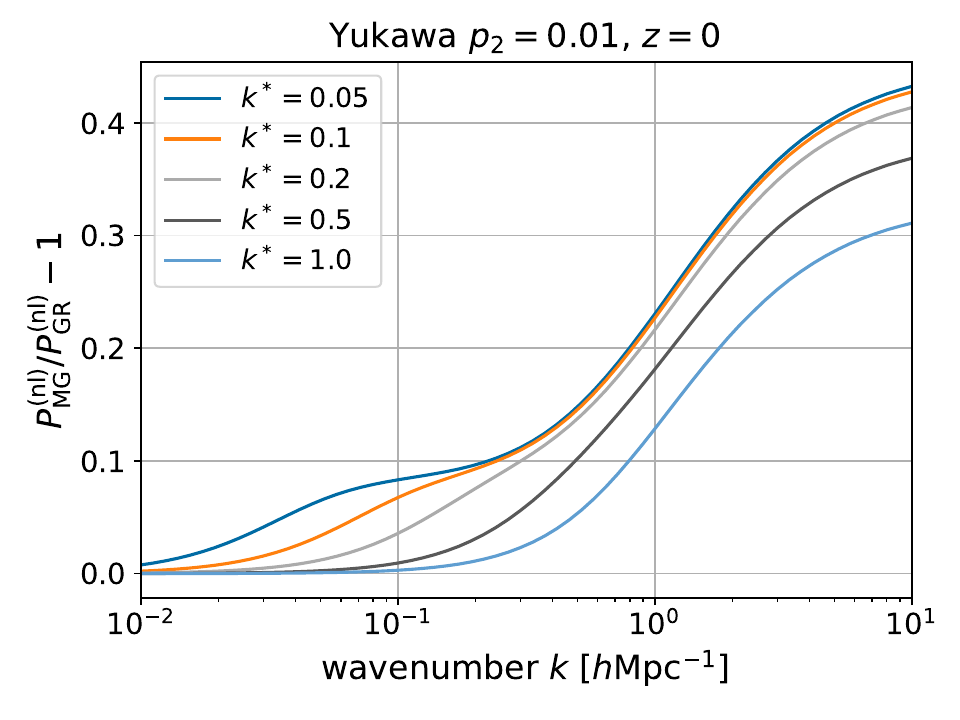}
		\caption{Relative difference between the non-linear power spectrum for GR and the Yukawa model for two different amplitudes $p_2$ and a range of different screening wave numbers, evaluated at $z=0$.}
		\label{fig:yukawa_nonlinear}
	\end{subfigure}
	\caption{}
	\label{fig:yukawa}
\end{figure}

As a last example we show the changes to the power spectrum for the Yukawa suppression model introduced in Sect.~\ref{sec:3}. In contrast to nDGP and k-Mouflage, this model suggests a modification of the power spectrum on small scales and no modification on the largest scales. Figure \ref{fig:yukawa} shows the results for two different amplitudes $p_2$ and a range of different screening wave numbers $k^*$. The relative change to the linear spectrum (see Fig.~\ref{fig:yukawa_linear}) reflects the shape of the gravitational coupling. For smaller $k^*$, the change sets in on larger scales, but all changes lead to the same constant relative change for large $k$, i.e.\ on small scales. Gravitational interaction increases the overall change, and smaller $k^*$ now leads to larger changes even for large $k$, see Fig.~\ref{fig:yukawa_nonlinear}. The authors of \cite{Hassani.2020} choose the significantly larger amplitude $p_2=1/3$ for the Yukawa suppression model. We inserted this amplitude as well, but ended up with heavily overestimated power on non-linear scales w.r.t.\ the $N$-body simulations. The reason is that \cite{Hassani.2020} use a time-dependent screening scale $k^*$, which effectively reduces the gravitational interaction strength at earlier cosmic times (see Eq.~(10) in \cite{Hassani.2020}). In order to come up with reasonable changes to the power spectrum, we therefore have to lower the amplitude $p_2$. We achieve good qualitative agreement with the results of \cite[Figure 1]{Hassani.2020}, e.g.\ more power on non-linear scales and a characteristic bump in the intermediate regime depending on the strength of the modification.

\section{Functional Taylor expansion of the non-linear density-fluctuation power spectrum}
\label{sec:5}

We now extend the functional Taylor expansion for describing non-linear effects of modified gravity theories using KFT, first described in \cite{Oestreicher.2023}, to the theories with screening effects discussed in this paper. This allows us to study the general effect modifying the gravitational theory has on the power spectrum irrespective of the precise theory of gravity chosen. The theory of gravity enters into the KFT formalism through the gravitational coupling strength, quantified by the gravitational constant $G$, and the expansion of the background space-time, quantified by the expansion function $E$. Any viable modified gravity (MG) theory will have to imply only small corrections to GR in order to agree with the multitude of observational constraints. This should allow to express the non-linear density-fluctuation power spectrum in an MG theory by a functional, first-order Taylor expansion around its form predicted by GR. In the models considered here, all that changes is the gravitational coupling strength both as a function of time and scale. The functional Taylor expansion thus takes the form
\begin{align}
	P_\delta^{(\mathrm{nl})}[G+\delta G](k, a) \approx
	P_\delta^{(\mathrm{nl})}[G](k,a) +
	\int_0^{\infty}\D \alpha\,\int_{a_{\mathrm{ini}}}^a \D x\,
	\frac{\delta P_\delta^{(\mathrm{nl})}[G](k,a)}{\delta G(\alpha,x)} \Delta G(\alpha,x)\;,
	\label{eq:Taylor_exp}
\end{align}
where 
\begin{align}
	\Delta G= G_\mathrm{MG}-G_\mathrm{GR}\;.
\end{align}
The functional derivatives quantify by how much the power spectrum, evaluated at wave number $k$ and scale factor $a$, reacts to changes in $G$ introduced at wave number $\alpha$ and scale factor $x$. Since changes can happen at any time preceding $a$ and any wave number, we integrate over $x$ and $\alpha$.

We emphasize here that the functional derivatives are evaluated in the GR limit and are therefore the same for all MG theories. This allows us to gain some general insight into how power spectra are modified, irrespective of the specific MG theory. 

The power spectrum depends on $G$ in three ways: directly through the interaction term, indirectly through the growth factor $D_+$, and further through the KFT time coordinate $t$. The change to the growth factor and the time coordinate have to be considered separately here because the growth factor gains acquires a $k$-dependence that has to be considered when calculating the linear power spectrum, while the time coordinate has to remain independent of $k$, as discussed in Sect.~\ref{sec:4}. 

Equation \eqref{eq:Taylor_exp} contains the functional derivative of the non-linear power spectrum with respect to $G$. Using the product rule,
\begin{align}
	\frac{\delta P_{\delta}^\mathrm{(nl)} (k,a)}{\delta G(\alpha, x)} =e^{\langle S_\mathrm{I} \rangle (k,a)}\frac{\delta P_\delta^\mathrm{(lin)}(k,a)}{\delta G(\alpha,x)}+ P_\delta^\mathrm{(nl)}(k,a) \frac{\delta \langle S_\mathrm{I} \rangle (k,a)}{\delta G(\alpha, x)}\;, 
\end{align}
it separates into two parts, a derivative of the linear power spectrum and a derivative of the mean interaction term. Finding these derivatives is the goal in the next two subsections.

\subsection{Functional derivative of the linear power spectrum}

The derivative of the linear power spectrum with respect to $G$ is given by
\begin{equation}
	\frac{\delta P_\delta^\mathrm{(lin)}(k,a)}{\delta G(\alpha,x)}=\frac{\delta}{\delta G(\alpha,x)}\left(D_+^2(k,a)P_\delta^\mathrm{(i)}(k)\right)=2P_\delta^\mathrm{(lin)}(k,a)\frac{\delta\ln D_+(k,a)}{\delta G(\alpha,x)}\;.
\end{equation}
The functional derivative of the growth factor appearing here, evaluated for GR, is calculated in Appx.~\ref{appendix:A1} and given by 
\begin{equation}
	\frac{\delta D_+(k,a)}{\delta G(\alpha,x)}\Bigg\vert_\mathrm{GR} = \delta_\mathrm{D}(k-\alpha)\frac{\delta D_+(a)}{\delta G(x)}\;, 
\end{equation}
where
\begin{equation}
	\frac{\delta D_+(a)}{\delta G(x)} = 
	\Theta(a-x)\,D_+(a)f_G(x)\int_x^a\frac{\D y}{D_+^2(y)y^3E(y)}
\end{equation}
with 
\begin{equation}
	f_G(x) = \frac{3}{2}\frac{\Omega_\mathrm{m}(x)}{G}D_+^2(x)xE(x)\;.
\end{equation}
This result agrees with what was found in \cite{Oestreicher.2023} up to a delta-function. The delta-function sets $\alpha=k$ in the integration in Eq.~\eqref{eq:Taylor_exp}, hence the linear spectrum changes its shape directly proportional to $\Delta G$ at first order in the Taylor expansion. The amplitude of the power spectrum changes in proportion to $\delta D_+(a)/\delta G(x)$. It was demonstrated in \cite{Oestreicher.2023} that this derivative is always positive and the larger the earlier the perturbation is applied. Since the derivative of the linear power spectrum is proportional to this derivative, a positive change in $G$ will always lead to an enhanced amplitude of the linear power spectrum. This confirms the intuition that stronger gravitational coupling leads to enhanced structure growth. 

\subsection{Functional derivative of the interaction term}
We calculate the functional derivative of the interaction term in Appx.~\ref{appendix:A2} it is given by
\begin{equation}
	\frac{\delta\langle S_\mathrm{I}\rangle (k,a)}{\delta G(\alpha, x)}\Bigg\vert_\mathrm{GR} = \frac{\delta\langle S_\mathrm{I}\rangle (k,a)}{\delta G(\alpha, x)}\Bigg\vert_\mathrm{GR}^{(1)}  + \frac{\delta\langle S_\mathrm{I}\rangle (k,a)}{\delta G(\alpha, x)}\Bigg\vert_\mathrm{GR}^{(2)}\;,
\end{equation}
where
\begin{align}
	\frac{\delta\langle S_\mathrm{I}\rangle (k,a)}{\delta G(\alpha, x)}\Bigg\vert_\mathrm{GR}^{(1)} =&\; 3\Theta(a-x)\frac{g_\mathrm{H}(a,x)}{x^2E(x)}\frac{1}{(2\pi)^2}\frac{k\alpha^3}{k_0^2 + \alpha^2} I_\mu(k,\alpha,x) \nonumber \\
	& +6 \int_{a_\mathrm{min}}^a\D a'\,\frac{g_\mathrm{H}(a,a')}{a'^2E} \frac{\alpha^2}{(2\pi)^2} \bar P_{\delta}^{(\mathrm{lin})}(\alpha,a')\frac{\delta\ln D_+(a')}{\delta G(x)}G J(\alpha/k,k_0/k)\;,
	\label{eq:dSI_dG1}
\end{align}
\begin{align}
	\frac{\delta\langle S_\mathrm{I}\rangle (k,a)}{\delta G( x)}\Bigg\vert_\mathrm{GR}^{(2)} =& -3 \int_{a_\mathrm{min}}^a\D a'\,\frac{g_\mathrm{H}(a,a')}{a'^2E}\frac{\delta D_+(a')}{\delta G(x)}G \nonumber \\ &\int_0^\infty\frac{\D k'\,k'^2}{(2\pi)^2} P_{\delta}^{(\mathrm{lin})}(k',a')(1+Q_\mathrm{D})^{-2}k'^2\lambda^2\left(
	\frac{2}{t}-\frac{1}{\sqrt{t\tau}+t}\right)J(k'/k,k_0/k)\;.
	\label{eq:dSI_dG2}
\end{align}
The second expression results entirely from the derivative of the damping term and is independent of the perturbance wave number $\alpha$. This derivative provides insight into the effect of modified gravitational coupling on the interaction term. We plot the two terms in Eqs.\ \eqref{eq:dSI_dG1} and \eqref{eq:dSI_dG2} separately, both as a function of the perturbance wave number $\alpha$ and the perturbance scale factor $x$. We show a range of fixed wave numbers $k=0.1,1,5,10$ and evaluate at scale factor $a=1$.

\begin{figure}[tbp]
	\centering
	\includegraphics[width=0.5\linewidth]{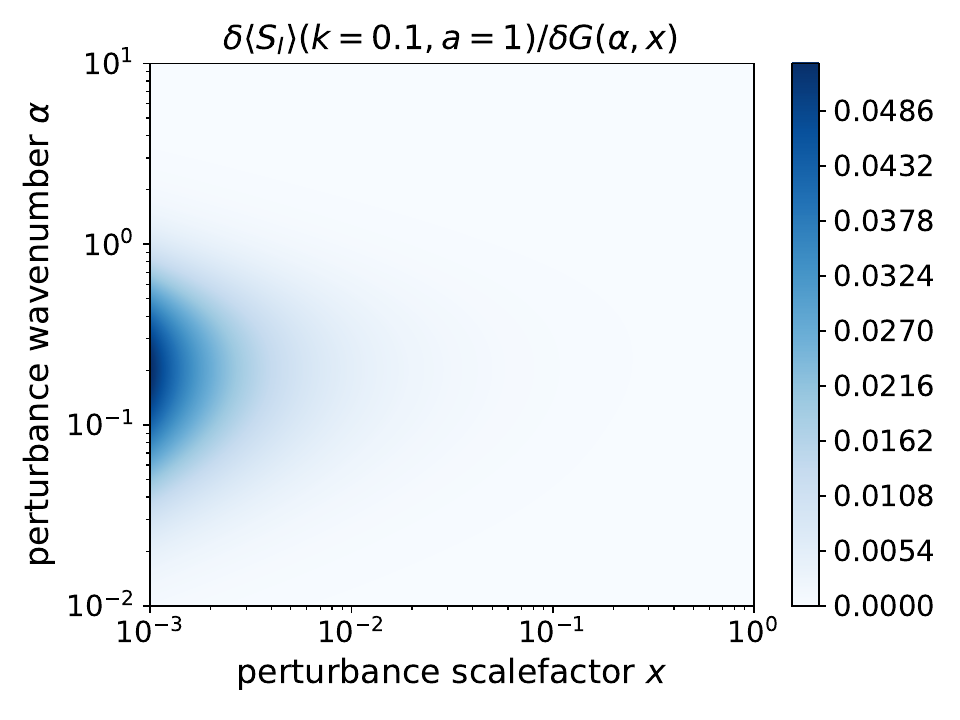}%
	\includegraphics[width=0.5\linewidth]{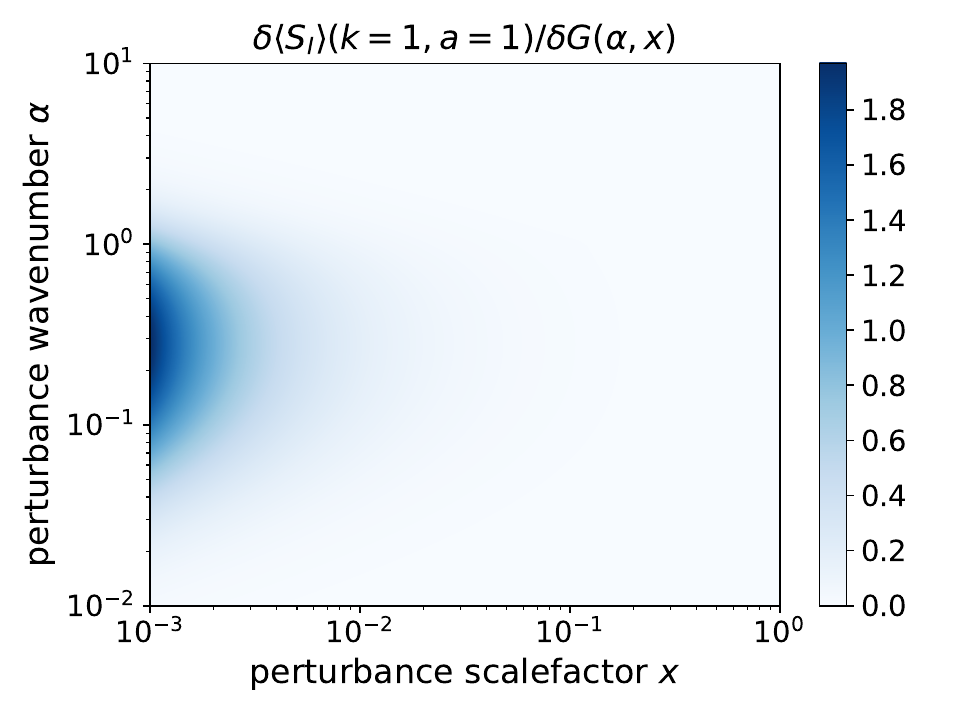}
	\includegraphics[width=0.5\linewidth]{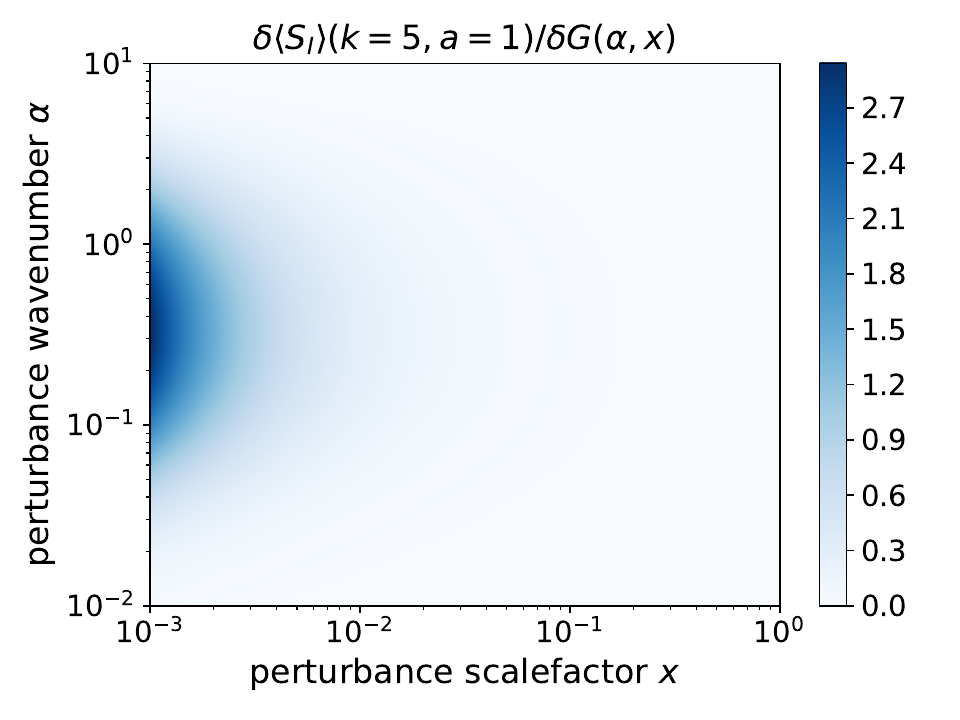}%
	\includegraphics[width=0.5\linewidth]{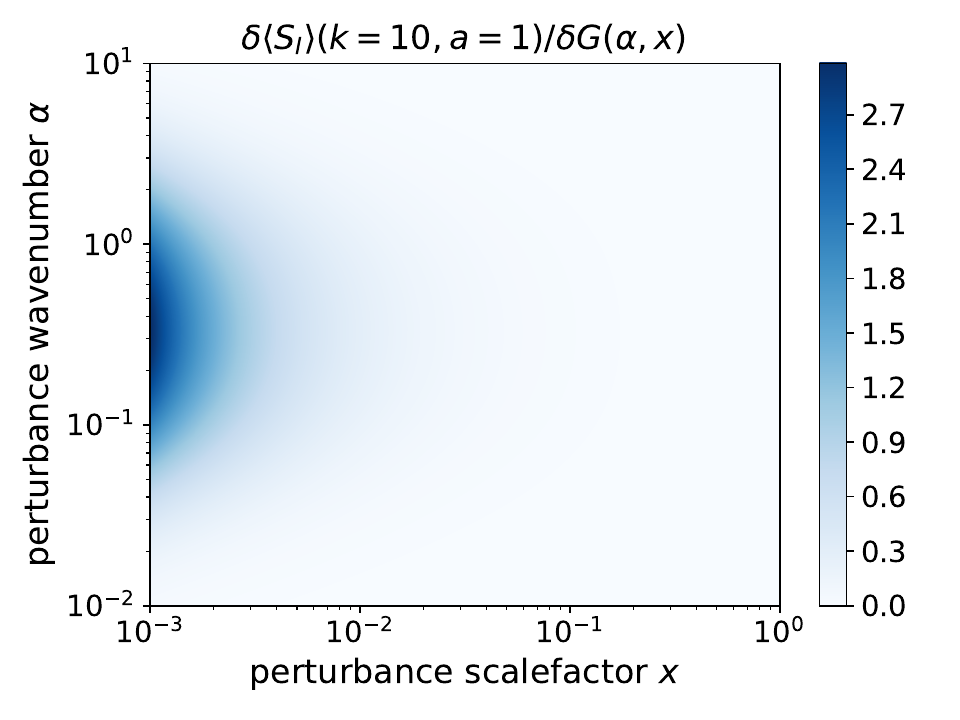}
	\caption{Part 1 of the functional derivative of the interaction term $\langle S_\mathrm{I} \rangle (k,a)$ w.r.t. $G(\alpha, x)$ with the vertical axis being the perturbance wave number $\alpha$, and the horizontal axis being the perturbance scale factor $x$. 
	}
	\label{fig:dSI_dG1}
\end{figure}

Figure~\ref{fig:dSI_dG1} shows the first term, i.e.\ Eq.~\eqref{eq:dSI_dG1}. We observe that the largest change occurs when the perturbation is introduced between $\alpha=0.1$ and $\alpha=1$, i.e.\ around the transition from linear to non-linear structure growth. This behaviour is generic. The larger the wavenumber $k$ is where the interaction term is evaluated, the larger is the effect of the perturbation. If we evaluate at a very small $k$ (top right plot), there is almost no change. This is because the spectrum evolves linearly on these scales, and $\langle S_\mathrm{I}\rangle$ is very nearly zero. The derivative is larger for smaller $x$ since earlier perturbations have more time to act and change structure growth. Lastly, the derivative is always positive, indicating that enhanced gravitational coupling leads to enhanced structure growth. In general, perturbations introduced at early times and around the scale where non-linear structure growth sets in effectuate the largest changes to the power spectrum and affect small scales (large $k$) the most.

\begin{figure}[tbp]
	\centering
	\includegraphics[width=0.5\linewidth]{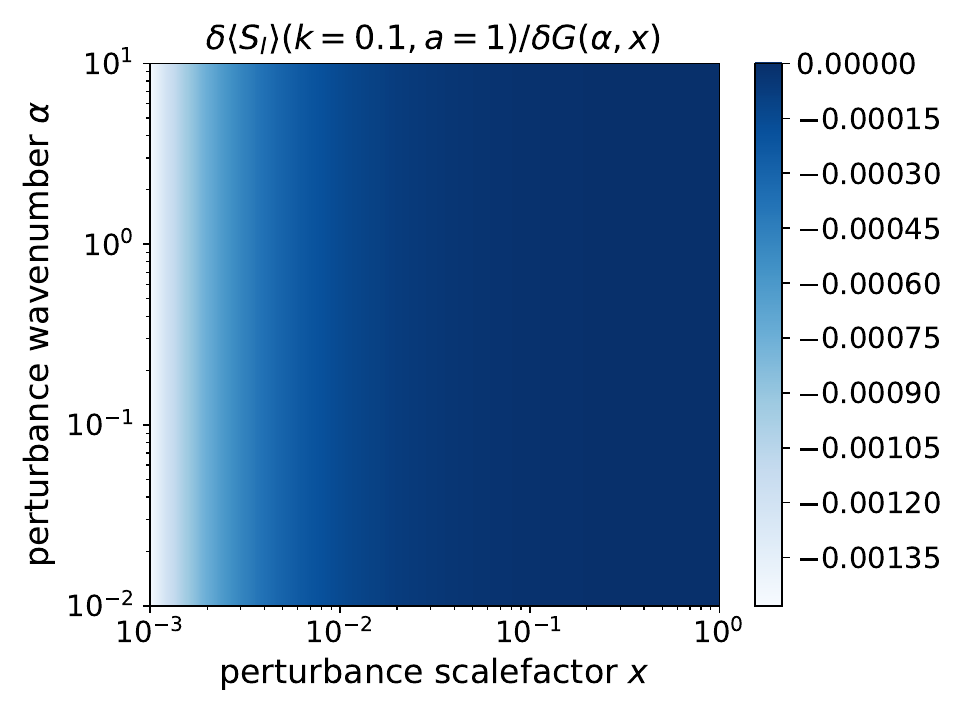}%
	\includegraphics[width=0.5\linewidth]{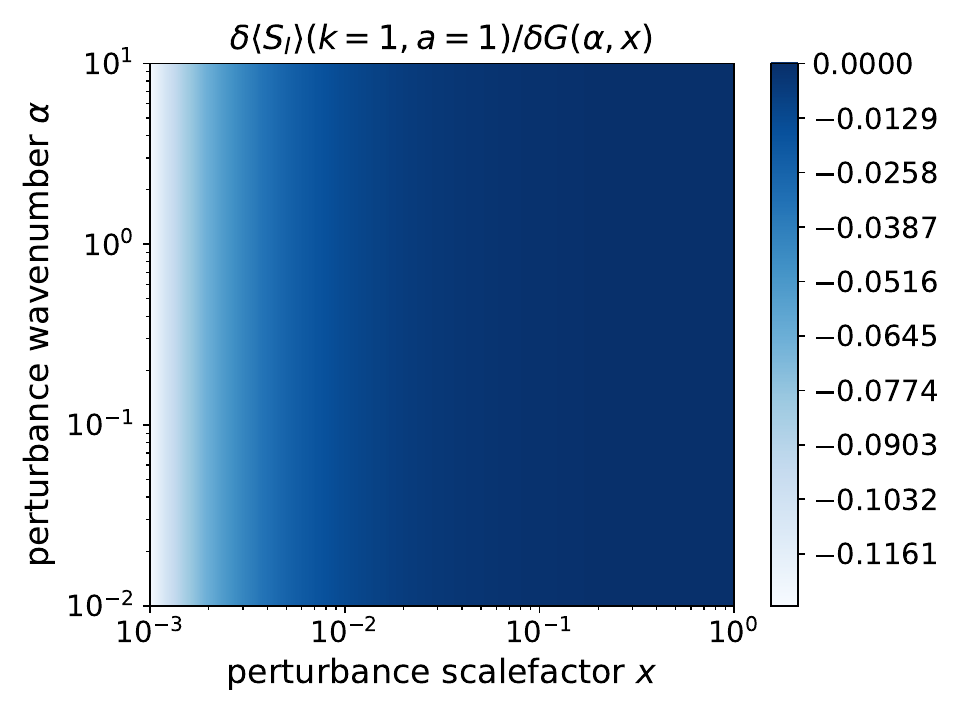}
	\includegraphics[width=0.5\linewidth]{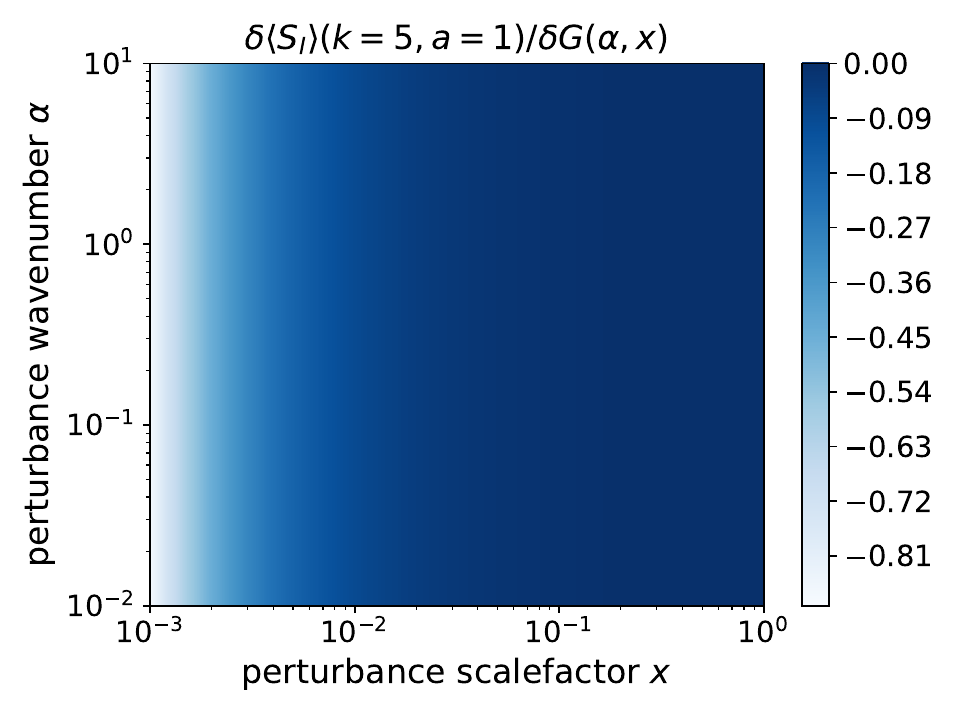}%
	\includegraphics[width=0.5\linewidth]{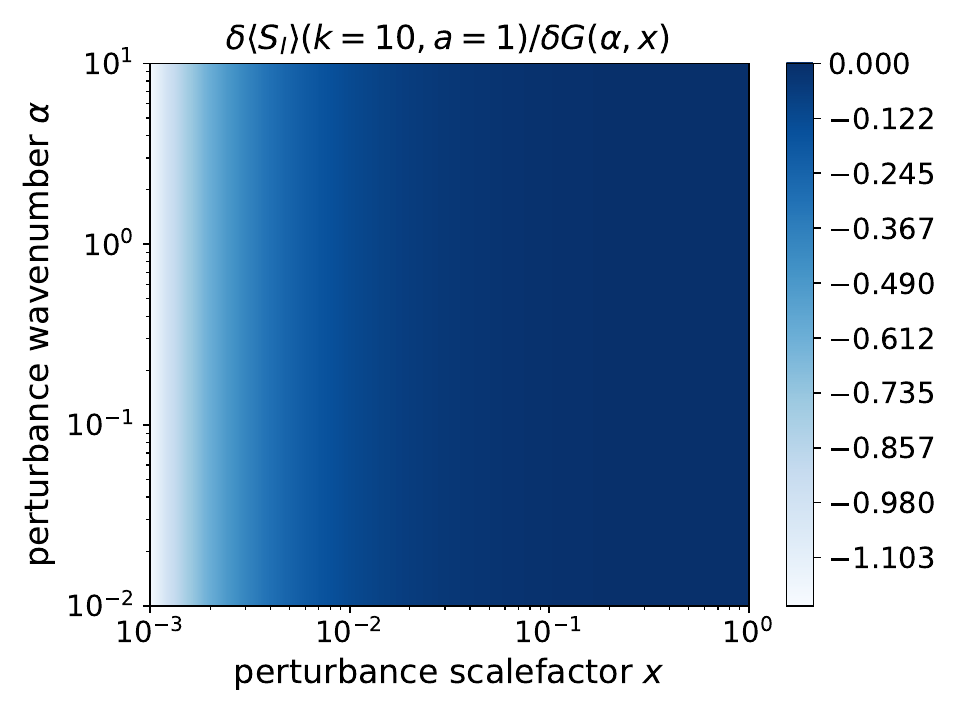}
	\caption{Part 2 of the functional derivative of the interaction term $\langle S_\mathrm{I} \rangle (k,a)$ with respect to $G(\alpha, x)$ with the vertical axis being the perturbance wave number $\alpha$, and the horizontal axis being the perturbance scale factor $x$. 
	}
	\label{fig:dSI_dG2}
\end{figure}

As Fig.~\ref{fig:dSI_dG2} shows, the second term in Eq.~\eqref{eq:dSI_dG2} does not depend on $\alpha$ and is always negative. The reason is that this term results purely from the damping term and changes only due to changes in our time coordinate $t$, which is independent of $k$. Overall changes to the spectrum should be positive, except for the inaccuracies related to the damping already discussed in Sect.~\ref{sec:4}. 

\subsection{Taylor-expanded non-linear spectra}

\begin{figure}[tbp]
	\centering
	\includegraphics[width=0.5\linewidth]{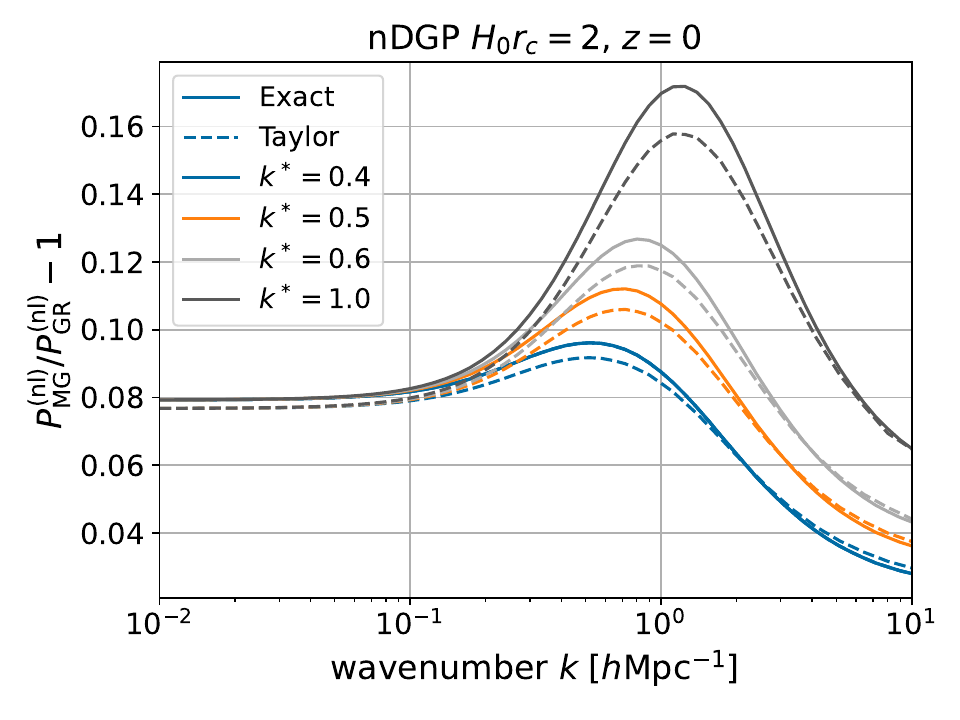}%
	\includegraphics[width=0.5\linewidth]{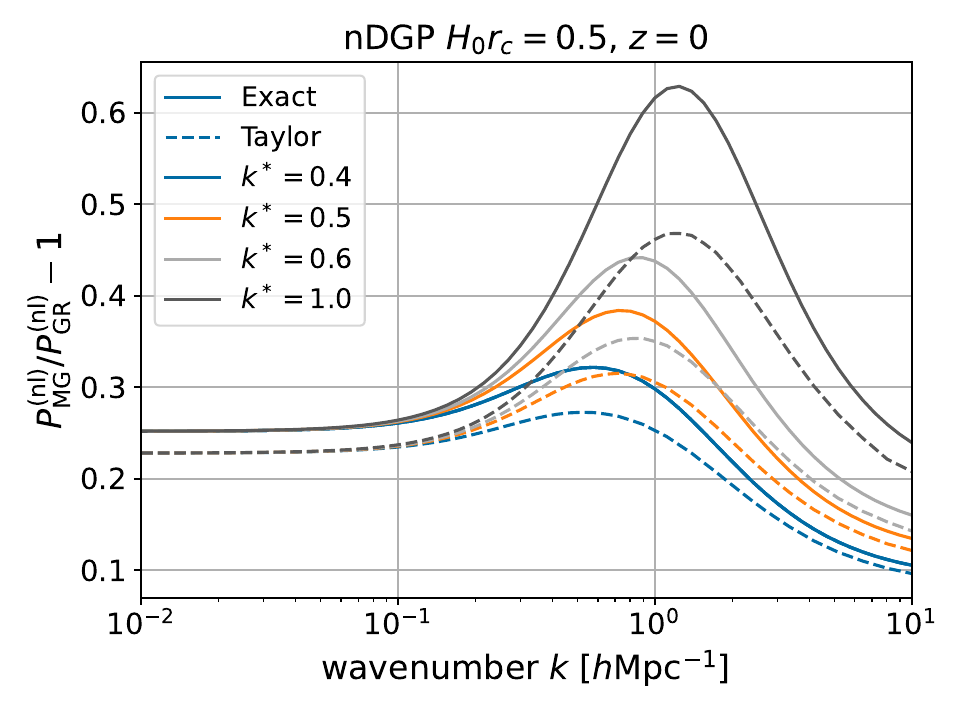}
	\caption{Relative difference between the non-linear power spectrum for GR and the nDGP model for two different crossover scales $r_\mathrm{c}$ and a range of different screening wave numbers, normalized at $z\simeq 1000$. For both the exact approach of Sect.~\ref{sec:4} and the Taylor expansion of Sect.~\ref{sec:5}.}
	\label{fig:ndgp_nonlinear_combined}
\end{figure}

\begin{figure}[tbp]
	\centering
	\includegraphics[width=0.5\linewidth]{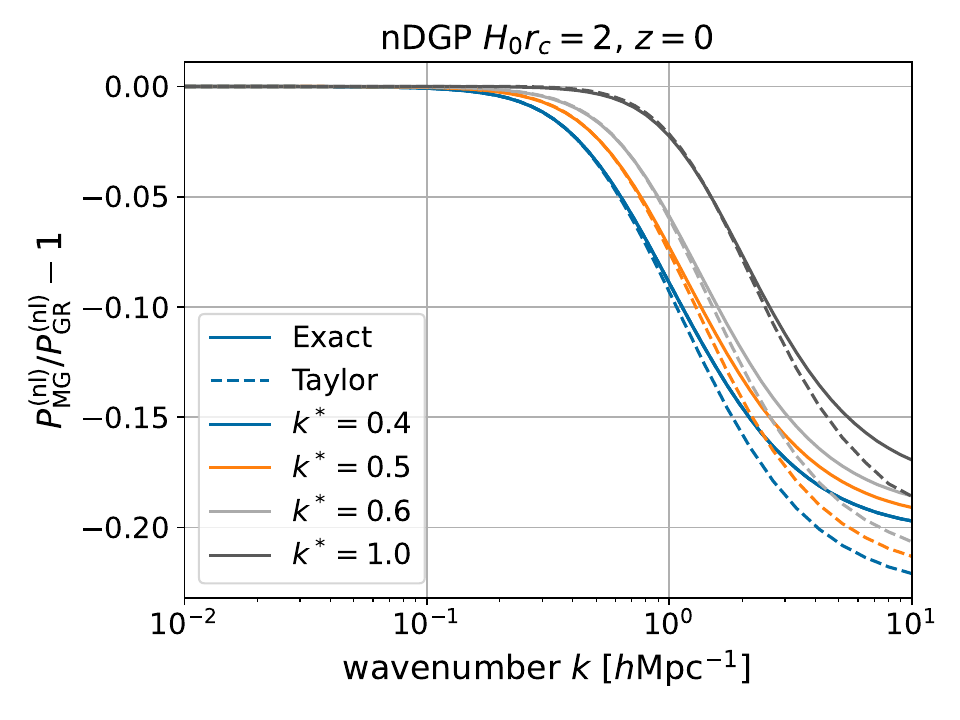}%
	\includegraphics[width=0.5\linewidth]{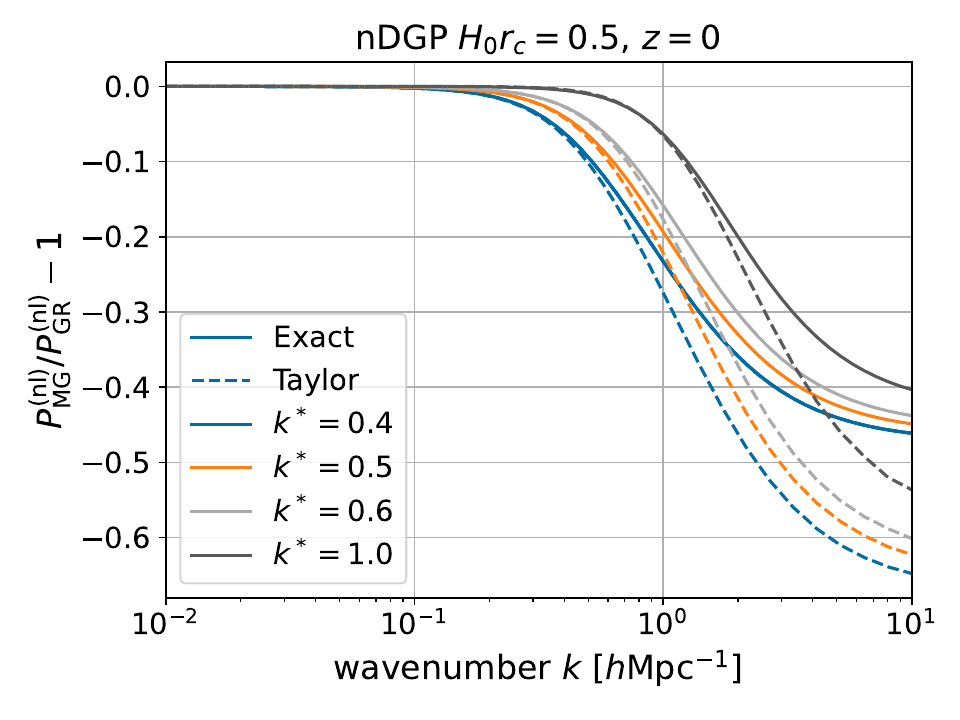}
	\caption{Relative difference between the non-linear power spectrum for GR and the nDGP model for two different crossover scales $r_\mathrm{c}$ and a range of different screening wave numbers, normalized at z=0. For both the exact approach of Sect.~\ref{sec:4} and the Taylor expansion of Sect.~\ref{sec:5}.}
	\label{fig:ndgp_nonlinear_an1_combined}
\end{figure}

Having calculated the functional derivatives, we can now evaluate power spectra using the functional Taylor expansion. The spectrum for a modified gravity theory reads
\begin{align}
	P_{\delta,\mathrm{MG}}^\mathrm{(nl)}(k,a) &\approx 
	P_{\delta,\mathrm{GR}}^\mathrm{(nl)}(k,a) + \int_{a_\mathrm{min}}^a\D x\,\int_0^{\infty}\D \alpha\,
	\Bigg( e^{\langle S_\mathrm{I} \rangle (k,a)}\frac{\delta P_\delta^\mathrm{(lin)}(k,a)}{\delta G(\alpha,x)}\Bigg\vert_\mathrm{GR} \nonumber \\
	&+ P_\delta^\mathrm{(nl)}(k,a) \frac{\delta \langle S_\mathrm{I} \rangle (k,a)}{\delta G(\alpha, x)}\Bigg\vert_\mathrm{GR}^{(1)} \Bigg)\Delta G(\alpha,x) + \int_{a_\mathrm{min}}^a\D x\, P_\delta^\mathrm{(nl)}(k,a) \frac{\delta \langle S_\mathrm{I} \rangle (k,a)}{\delta G(\alpha, x)}\Bigg\vert_\mathrm{GR}^{(2)} \Delta G^\mathrm{(lin)}(x) \;,
\end{align}
where we integrate the second term only over $x$, since it results purely from changes to the time coordinate and does therefore not depend on $k$. With the large-scale limit of the MG gravitational coupling defined earlier, $\Delta G^\mathrm{(lin)} = G_\mathrm{MG}^\mathrm{(lin)}-G_\mathrm{GR}$. Inserting the derivative of the linear power spectrum and integrating over the appearing delta distribution, we find
\begin{align}
	P_{\delta,\mathrm{MG}}^\mathrm{(nl)}(k,a) &\approx 
	P_{\delta,\mathrm{GR}}^\mathrm{(nl)}(k,a) \Bigg( 1+ \int_{a_\mathrm{min}}^a\D x\,
	2\frac{\delta\ln D_+(a)}{\delta G(x)}\Bigg\vert_\mathrm{GR}\Delta G(k,x) \nonumber \\
	&+ \int_{a_\mathrm{min}}^a\D x\,\int_0^{\infty}\D \alpha\,\frac{\delta \langle S_\mathrm{I} \rangle (k,a)}{\delta G(\alpha, x)}\Bigg\vert_\mathrm{GR}^{(1)} \Delta G(\alpha,x) + \int_{a_\mathrm{min}}^a\D x\, \frac{\delta \langle S_\mathrm{I} \rangle (k,a)}{\delta G(\alpha, x)}\Bigg\vert_\mathrm{GR}^{(2)} \Delta G^\mathrm{(lin)}(x)\Bigg) \;.
\end{align}
The amplitude changes discussed earlier can be taken into account by derivatives calculated in \cite{Oestreicher.2023}. One finds 
\begin{align}
	P_{\delta,\mathrm{MG}}^\mathrm{(nl)}(k,a) &\approx 
	P_{\delta,\mathrm{GR}}^\mathrm{(nl)}(k,a) \Bigg( 1+ \int_{a_\mathrm{min}}^a\D x\,
	2\frac{\delta\ln D_+(a)}{\delta G(x)}\Bigg\vert_\mathrm{GR}\Delta G(k,x) \nonumber \\
	&+ \int_{a_\mathrm{min}}^a\D x\,\int_0^{\infty}\D \alpha\,\frac{\delta \langle S_\mathrm{I} \rangle (k,a)}{\delta G(\alpha, x)}\Bigg\vert_\mathrm{GR}^{(1)} \Delta G(\alpha,x) + \int_{a_\mathrm{min}}^a\D x\, \frac{\delta \langle S_\mathrm{I} \rangle (k,a)}{\delta G(\alpha, x)}\Bigg\vert_\mathrm{GR}^{(2)} \Delta G^\mathrm{(lin)}(x) \nonumber \\
	&-\int_{a_\mathrm{min}}^a\D x\, 2\frac{\partial P_{\delta,\mathrm{GR}}^\mathrm{(nl)}(k,a)}{\partial\mathcal{A}}
	\frac{\delta\ln D_+(a)}{\delta G(x)}\Bigg\vert_\mathrm{GR}\Delta G^\mathrm{(lin)}(x)\Bigg) \;,
\end{align}
where the derivative of the non-linear power spectrum w.r.t.\ $\mathcal{A}$ evaluated in GR is 
\begin{equation}
	\frac{\partial P^\mathrm{(nl)}_{\delta,\mathrm{GR}}(k,a)}{\partial\mathcal{A}} =
	\frac{P^\mathrm{(nl)}_{\delta,\mathrm{GR}}(k,a)}{\mathcal{A}}+
	P^\mathrm{(nl)}_{\delta,\mathrm{GR}}(k,a)\,
	\frac{\partial\langle S_\mathrm{I}\rangle(k,a)}{\partial\mathcal{A}}\Bigg\vert_\mathrm{GR}
	\label{eq:36}
\end{equation}
with
\begin{equation}
	\frac{\partial\langle S_\mathrm{I}\rangle(k,a)}{\partial\mathcal{A}}\Bigg\vert_\mathrm{GR} =
	3\int_{a_\mathrm{min}}^a\D a'\,\frac{g_\mathrm{H}(a,a')}{a'^2E}D_+^2G
	\frac{\partial\sigma_J^2\left(k, t'\right)}{\partial\mathcal{A}}\;,
	\label{eq:37}
\end{equation}
and
\begin{equation}
	\frac{\partial\sigma_J^2(k,a)}{\partial\mathcal{A}} =
	\frac{1}{(2\pi)^2}\int_0^\infty\,\D y\, y^2
	\left(1+\mathcal{A}Q_\mathrm{D}\right)^{-1}
	\mathcal{A}P^\mathrm{(i)}(y)\left[
	\frac{1}{\mathcal{A}}-Q_\mathrm{D}\left(1+\mathcal{A}Q_\mathrm{D}\right)^{-1}
	\right]J_\mu(y/k,y_0/k)\;.
	\label{eq:38}
\end{equation}
We evaluate the Taylor-expanded, non-linear spectra and compare them to the non-linear spectra exactly calculated earlier for the case of nDGP gravity in Figs.~\ref{fig:ndgp_nonlinear_combined} and \ref{fig:ndgp_nonlinear_an1_combined}. The Taylor expansion is very accurate for small deviations from the GR result. For relative deviations of the non-linear power spectrum of $\lesssim 15\%$, the amplitude of the Taylor expansion agrees within $1\%$ relative accuracy with the exact result. For larger relative deviations the amplitude of the Taylor expansion falls below the exact result, but in all cases, the characteristic shape of the deviation between the two different power spectra is well reproduced. This demonstrates that the general insight gained from the Taylor expansion and in particular the shape of the functional derivatives corresponds well to the exact solutions. Better agreement for larger relative deviations could be achieved by including higher orders in the Taylor expansion, but as our main interest in the Taylor expansion was to gain insight into how the power spectrum responds to modifications of the gravitational theory, and as the shape is already very well reproduced for all theories, this seems unnecessary. Changes to the power spectrum also have to remain small in order to agree with current observations, further ensuring that the first order Taylor expansion and the insights gained from it remain valid.  The results are equally accurate for the other theories studied in this paper, so we do not show them here.

\section{Conclusions}

Cosmic structure formation can analytically be described by Kinetic Field Theory (KFT). By construction, and in contrast to other analytic approaches, KFT avoids the shell-crossing problem and can therefore proceed into the non-linear regime without conceptual limitations. Furthermore, it does not use the density contrast as a perturbed quantity but the deviation of particle trajectories from fiducial free trajectories. Non-linear density fluctuations can thus be reached even in the free theory.

The interaction operator elevating KFT to an interacting theory admits a mean-field approximation. Within this approximation, the GR power spectrum of non-linear cosmic density fluctuations can be analytically reproduced with percent level accuracy to wave numbers $k\lesssim10\,h\,\mathrm{Mpc}^{-1}$ at redshift zero. The corresponding mathematical expression multiplies the linearly evolved density-fluctuation power spectrum by an exponential factor containing an averaged mean-field interaction term and is thus easy to evaluate. It depends on the evolution of the background space-time only via the expansion function and on the gravity theory only via the gravitational coupling strength. 

This fact allows us to study effects that a modified gravity theory would have on the non-linear density-fluctuation power spectrum by simply changing the expansion function and gravitational coupling strength. Furthermore, changes to the power spectrum are expected to be small compared to the standard case. Otherwise the cosmological standard model could not be as successful as it is. This suggests a Taylor expansion of the non-linear density-fluctuation power spectrum w.r.t.\ the expansion function and the gravitational coupling strength. Since the Taylor expansion is taken around the GR case, the functional derivatives only need to be evaluated once under the assumptions of the cosmological standard model. This Taylor expansion and in particular the shape of the functional derivatives provide generic insight into the changes to the power spectrum under a modification of the gravity theory, irrespective of its specifics. 

We have shown in earlier work \cite{Oestreicher.2023} how the power spectrum would change under generic time-dependent changes to the expansion function and gravitational coupling, but we had not yet discussed such modified gravity theories with a screening mechanism. In this paper, we have filled this gap.

We have demonstrated how these screening mechanisms can be included into the KFT mean-field formalism by introducing a scale dependent effective gravitational coupling, allowing to calculate exact (within the mean-field approximation) power spectra for any specific theory of gravity. We have also studied a functional Taylor expansion of the power spectrum around general relativity, allowing us to understand how the non-linear power spectrum changes, quite independently of the specific theory of gravity chosen. For small deviations from the cosmological standard model, the functional Taylor expansion turns out to be very accurate and in good agreement with the full mean-field results.

Adopting the parametrised model for the screening effects on the gravitational coupling strength suggested by Lombriser et al. \cite{Lombriser.2016}, we have evaluated the non-linear density-fluctuation power spectrum for three different screening models. This allowed us to compare our results to numeric results in the literature \cite{Cataneo.2019, Hassani.2020}, showing good qualitative agreement in all cases, and percent-level agreement with the precise numeric simulations by \cite{Cataneo.2019} in the specific case of nDGP gravity for wavenumbers $k\lesssim 2\,h\,\text{Mpc}^{-1}$. 

The analytic framework of KFT and resulting quick calculation of power spectra now allows a systematic scanning of broad classes of modified gravity theories, with and without screening, without having to rely on costly numerical simulations. Large theory and parameter spaces can be studied in a few hours on any personal laptop, and possibly interesting parameter spaces can be selected for further numerical evaluation if desired. Additionally the analytic approach provides insight into how any changes to the gravity theory affect the density-fluctuation power spectrum, without having to specify the modification of the gravitational theory. 

\acknowledgments
This work was funded in part by Deutsche Forschungsgemeinschaft (DFG, German Research Foundation) under Germany’s Excellence Strategy EXC-2181/1 - 390900948 (the Heidelberg STRUCTURES Cluster of Excellence). During the final stages of this project, AO was funded by VILLUM FONDEN, grant VIL53032. HS thanks DAAD-WISE for funding during the summer internship that started this project. EK is funded by the Deutsche Forschungsgemeinschaft (DFG, German Research Foundation) – 452923686. During the early stages of this work, JD was funded by a Research Leadership Award from the Leverhulme Trust awarded to Clare Burrage. We thank Lucas Lombriser and Farbod Hassani for helpful discussion and sharing their numeric results with us. We also thank Ricardo Waibel for helpful discussions and his contributions during the initial stages of the project.

\appendix 

\section{Functional derivatives}
\label{appendix:A}
\subsection{Functional derivative of the growth factor}
\label{appendix:A1}
The functional derivative of the linear growth factor $D_+$ with respect to $G$ can be obtained by taking a functional derivative of the linear growth equation \eqref{eq:lin_growth_eq}. It is
\begin{align}
	&\frac{\D^2}{\D a^2}\frac{\delta D_+(k,a)}{\delta G(\alpha,x)}+\left(
	\frac{3}{a}+\frac{E'}{E}
	\right)\frac{\D}{\D a}\frac{\delta D_+(k,a)}{\delta G(\alpha,x)}-
	\frac{3}{2}\frac{\Omega_\mathrm{m}G}{a^2G}\left(
	\frac{\delta D_+(k,a)}{\delta G(\alpha,x)}
	\right) \nonumber \\
	&= -\frac{3}{2}\frac{\Omega_\mathrm{m} D_+}{a^2G}\left(
	\frac{\delta G(k,a)}{\delta G(\alpha,x)}
	\right) \nonumber \\
	&=-\frac{3}{2}\frac{\Omega_\mathrm{m} D_+}{a^2G}\delta_\mathrm{D}(k-\alpha)\delta_\mathrm{D}(a-x)\;.
\end{align}
This differential equation has the same form as Eq.~\eqref{eq:lin_growth_eq}. This suggests the ansatz $\delta D_+/\delta G = C(a,x,\alpha,k)D_+(k,a)$, corresponding to a variation of constants. Solving this ansatz and enforcing causality leads to
\begin{equation}
	\frac{\delta D_+(k,a)}{\delta G(\alpha,x)} = \delta_\mathrm{D}(k-\alpha)
	\Theta(a-x)\,D_+(k,a)f_G(x)\int_x^a\frac{\D y}{D_+^2(k,y)y^3E(y)}
\end{equation}
with 
\begin{equation}
	f_G(x) = \frac{3}{2}\frac{\Omega_\mathrm{m}(x)}{G}D_+^2(k,x)xE(x)\;.
\end{equation}
We will only need to evaluate this result in the GR case, where $D_+$ does not depend on $k$. Therefore, this result agrees with what was found in \cite{Oestreicher.2023} up to a delta-function,
\begin{equation}
	\frac{\delta D_+(k,a)}{\delta G(\alpha,x)}\Bigg\vert_\mathrm{GR} = \delta_\mathrm{D}(k-\alpha)\frac{\delta D_+(a)}{\delta G(x)}\;, 
\end{equation}
where
\begin{equation}
	\frac{\delta D_+(a)}{\delta G(x)} = 
	\Theta(a-x)\,D_+(a)f_G(x)\int_x^a\frac{\D y}{D_+^2(y)y^3E(y)}
\end{equation}
with 
\begin{equation}
	f_G(x) = \frac{3}{2}\frac{\Omega_\mathrm{m}(x)}{G}D_+^2(x)xE(x)\;.
\end{equation}

\subsection{Functional derivative of the interaction term}
\label{appendix:A2}
For the derivative of the interaction term, we use the general expression in Eq.~\eqref{eq:SI} as a starting point. The product rule implies
\begin{align}
	\frac{\delta\langle S_\mathrm{I}\rangle (k,a)}{\delta G(\alpha, x)} =&\; 3 \int_{a_\mathrm{min}}^a\D a'\,\frac{g_\mathrm{H}(a,a')}{a'^2E} \vec k \cdot \int_{k'} \frac{\delta G(a',|\vec k -\vec k'|)}{\delta G(\alpha,x)}\frac{\vec k-\vec k'}{k_0^2+(\vec k-\vec k')^2}\bar P_{\delta}^{(\mathrm{lin})}(k',a') \nonumber \\
	& +3 \int_{a_\mathrm{min}}^a\D\,a'\frac{g_\mathrm{H}(a,a')}{a'^2E} \vec k \cdot \int_{k'} G(a',|\vec k -\vec k'|)\frac{\vec k-\vec k'}{k_0^2+(\vec k-\vec k')^2}\frac{\delta \bar P_{\delta}^{(\mathrm{lin})}(k',a')}{\delta G(\alpha,x)}\;.
\end{align}
We first evaluate the integral over $k$-space and find
\begin{align}
	\frac{\delta\langle S_\mathrm{I}\rangle (k,a)}{\delta G(\alpha, x)} =&\; 3 \int_{a_\mathrm{min}}^a\D a'\, \frac{g_\mathrm{H}(a,a')}{a'^2E}\int_{0}^{\infty}\frac{\D k'\,k'^2}{(2\pi)^2}\frac{kk'}{k_0^2 + k'^2}\frac{\delta G(a',k')}{\delta G(\alpha,x)}\nonumber \\ 
	&\int_{-1}^{1}\D\mu\, \mu \Bar{P}_\delta^\mathrm{(i)} \left(\sqrt{k^2+k'^2-2kk'\mu},a'\right) \nonumber \\
	& +3 \int_{a_\mathrm{min}}^a\D\,a'\frac{g_\mathrm{H}(a,a')}{a'^2E} \int_0^\infty \frac{\D k'\,k'^2}{(2\pi)^2} \frac{\delta \bar P_{\delta}^{(\mathrm{lin})}(k',a')}{\delta G(\alpha,x)} \nonumber \\ &\int_{-1}^1 \D\mu\,G\biggl(a',\sqrt{k^2+k'^2-2kk'\mu}\biggr)\,\frac{k^2-kk'\mu}{k_0^2+k^2+k'^2-2kk'\mu}\;,
\end{align}
where we used again in the first summand that the convolution commutes. The two integrals over $\mu$ can be evaluated numerically. We introduce the short hand notations  $I_\mu$ and $J_\mu$ to write
\begin{align}
	\label{eq:dSIdG}
	\frac{\delta\langle S_\mathrm{I}\rangle (k,a)}{\delta G(\alpha, x)} =&\; 3 \int_{a_\mathrm{min}}^a\D a'\, \frac{g_\mathrm{H}(a,a')}{a'^2E}\int_{0}^{\infty}\frac{\D k'\,k'^2}{(2\pi)^2}\frac{kk'}{k_0^2 + k'^2}\frac{\delta G(a',k')}{\delta G(\alpha,x)} I_\mu(k,k',a') \nonumber \\
	& +3 \int_{a_\mathrm{min}}^a\D a'\,\frac{g_\mathrm{H}(a,a')}{a'^2E} \int_0^\infty \frac{\D k'\,k'^2}{(2\pi)^2} \frac{\delta \bar P_{\delta}^{(\mathrm{lin})}(k',a')}{\delta G(\alpha,x)} J_\mu(k,k',a')\;.
\end{align}
When evaluated in GR, where $G$ depends neither on scale nor on time, the function $J_\mu$ can be integrated analytically; see \cite{Bartelmann.2021}. This gives
\begin{equation}
	J_\mu(\kappa,\kappa_0)\Big\vert_\mathrm{GR}=GJ(\kappa, \kappa_0)=G \left(1+\frac{1-\kappa^2-\kappa_0^2}{4\kappa}\ln\frac{\kappa_0^2+(1+\kappa)^2}{\kappa_0^2+(1+\kappa)^2}\right)\;, 
\end{equation}
where we introduced the variables $\kappa=k'/k$ and $\kappa_0=k_0/k$. The functional derivative of the damped linear power spectrum w.r.t.\ $G$ is
\begin{align}
	\label{eq:dPdampeddG}
	\frac{\delta \bar P_{\delta}^{(\mathrm{lin})}(k',a')}{\delta G(\alpha,x)}=&\;\frac{\delta}{\delta G(\alpha,x)}\left(\left(1+Q_\mathrm{D}\right)^{-1}
	D_+^2(k',a')P^\mathrm{(i)}_\delta(k')\right) \nonumber \\
	=&\;2\bar P_{\delta}^{(\mathrm{lin})}(k',a')\frac{\delta\ln D_+(k',a')}{\delta G(\alpha,x)}+P_{\delta}^{(\mathrm{lin})}(k',a')\frac{\partial}{\partial t}\left(1+Q_\mathrm{D}\right)^{-1}\frac{\delta t(a')}{\delta G(\alpha,x)} \nonumber \\
	=&\; 2\bar P_{\delta}^{(\mathrm{lin})}(k',a')\frac{\delta\ln D_+(k',a')}{\delta G(\alpha,x)} \nonumber \\ 
	&-P_{\delta}^{(\mathrm{lin})}(k',a')(1+Q_\mathrm{D})^{-2}k'^2\lambda^2\left(
	\frac{2}{t}-\frac{1}{\sqrt{t\tau}+t}\right)\frac{\delta D_+(a')}{\delta G(x)}\;.
\end{align}
Since the KFT time coordinate does not depend on $k$, its derivative here is reduced to the functional derivative of the growth factor without scale dependence. 

Inserting Eq.~\eqref{eq:dPdampeddG} into Eq.~\eqref{eq:dSIdG} and using that the functional derivative of $G(k',a'$) w.r.t.\ $G(\alpha,x)$ creates two delta-functions, we find
\begin{align}
	\frac{\delta\langle S_\mathrm{I}\rangle (k,a)}{\delta G(\alpha, x)}\Bigg\vert_\mathrm{GR} =&\; 3\Theta(a-x)\frac{g_\mathrm{H}(a,x)}{x^2E(x)}\frac{1}{(2\pi)^2}\frac{k\alpha^3}{k_0^2 + \alpha^2} I_\mu(k,\alpha,x) \nonumber \\
	& +6 \int_{a_\mathrm{min}}^a\D a'\,\frac{g_\mathrm{H}(a,a')}{a'^2E} \frac{\alpha^2}{(2\pi)^2} \bar P_{\delta}^{(\mathrm{lin})}(\alpha,a')\frac{\delta\ln D_+(a')}{\delta G(x)}G J(\alpha/k,k_0/k)\nonumber \\
	& -3 \int_{a_\mathrm{min}}^a\D a'\,\frac{g_\mathrm{H}(a,a')}{a'^2E}\frac{\delta D_+(a')}{\delta G(x)}G \nonumber \\ &\int_0^\infty\frac{\D k'\,k'^2}{(2\pi)^2} P_{\delta}^{(\mathrm{lin})}(k',a')(1+Q_\mathrm{D})^{-2}k'^2\lambda^2\left(
	\frac{2}{t}-\frac{1}{\sqrt{t\tau}+t}\right)J(k'/k,k_0/k)\;.
\end{align}
The third summand, resulting from the derivative of the time coordinate $t$, does not depend on the perturbance wave number $\alpha$. We split it from the other two terms and introduce the terms
\begin{align}
	\frac{\delta\langle S_\mathrm{I}\rangle (k,a)}{\delta G(\alpha, x)}\Bigg\vert_\mathrm{GR}^{(1)} =&\; 3\Theta(a-x)\frac{g_\mathrm{H}(a,x)}{x^2E(x)}\frac{1}{(2\pi)^2}\frac{k\alpha^3}{k_0^2 + \alpha^2} I_\mu(k,\alpha,x) \nonumber \\
	& +6 \int_{a_\mathrm{min}}^a\D a'\,\frac{g_\mathrm{H}(a,a')}{a'^2E} \frac{\alpha^2}{(2\pi)^2} \bar P_{\delta}^{(\mathrm{lin})}(\alpha,a')\frac{\delta\ln D_+(a')}{\delta G(x)}G J(\alpha/k,k_0/k)\;,
\end{align}
\begin{align}
	\frac{\delta\langle S_\mathrm{I}\rangle (k,a)}{\delta G( x)}\Bigg\vert_\mathrm{GR}^{(2)} =& -3 \int_{a_\mathrm{min}}^a\D a'\,\frac{g_\mathrm{H}(a,a')}{a'^2E}\frac{\delta D_+(a')}{\delta G(x)}G \nonumber \\ &\int_0^\infty\frac{\D k'\,k'^2}{(2\pi)^2} P_{\delta}^{(\mathrm{lin})}(k',a')(1+Q_\mathrm{D})^{-2}k'^2\lambda^2\left(
	\frac{2}{t}-\frac{1}{\sqrt{t\tau}+t}\right)J(k'/k,k_0/k)\;.
\end{align}

\bibliography{main}
\end{document}